\documentclass[symmetry,article,accept,moreauthors,12pt,a4paper]{mdpi}
\lastpage{x}
\doinum{10.3390/------}
\pubvolume{xx}
\pubyear{2019}
\history{Received: xx / Accepted: xx / Published: xx}

\usepackage{amssymb}

\Title{Nonextensive quasiparticle description of QCD matter}

\Author{Jacek Ro\.zynek, Grzegorz Wilk*}

\address{%
National Centre for Nuclear Research,~Department of Fundamental Research, Ho\.za 69, 00-681; Warsaw, Poland}

\corres{grzegorz.wilk@ncbj.gov.pl, +48-22-621 60 85}

\abstract{The dynamics of QCD matter is often described using effective mean field (MF) models based on Boltzmann-Gibbs (BG) extensive statistics. However, such matter is normally produced in small packets and in violent collisions where the usual conditions justifying the use of BG statistics are not fulfilled and the systems produced are not extensive. This can be accounted for either by enriching the original dynamics or by replacing the BG statistics by its nonextensive counterpart described by a nonextensivity parameter $q\neq 1$ (for $q \to 1$ one returns to the extensive situation). In this work we investigate the interplay between the effects of dynamics and nonextensivity.  Since the complexity of the nonextensive MF models prevents their simple visualization, we instead use some simple quasi-particle description of QCD matter in which the interaction is modelled phenomenologically by some effective fugacities, $z$. Embedding such a model in a nonextensive environment allows for a well-defined separation of the dynamics (represented by $z$) and the nonextensivity (represented by $q$) and a better understanding of their relationship.
}



\begin{document}

\section{Introduction}
\label{sec:Introduction}

Dense hadronic matter is usually described using relativistic mean field (MF) theory models (like, for example, the Walecka model for nucleons \cite{W,W1,W2} or the Nambu--Jona-Lasinio model (NJL) for quarks \cite{NJL,NJL1,NJL2,NJL3}). All of them use the Boltzmann-Gibbs (BG) statistics, which means that they assume a homogeneous and infinite heat bath and in their original versions they do not account for any intrinsic fluctuations or long range correlations. However, this kind of matter is typically produced in violent collision processes and in rather small packets, which rapidly evolve in a highly nonhomogeneous way and whose spatial configurations (like the correlations between quarks located in different nucleons in NJL models) remain far from being uniform (in fact, there is no global equilibrium established, cf. \cite{d1,d2,d4} and references therein). As a result, some quantities become non-extensive and develop power-law tailed rather than exponential distrubutions, making application of the usual BG statistics questionable (cf., \cite{WW1,WW3} and references therein). The remedy is either to supplement the BG statistics by some additional  dynamical input or, when it is not known, to use some form of nonextensive statistics generalizing the BG one, for example Tsallis statistics \cite{T,T2}. The latter is characterized by a nonextensivity parameter $q\neq 1$ (for $q=1$ one recovers the usual BG statistics). In fact, such an approach has already been investigated some time ago and the $q$-versions of essentially all types of MF models were formulated (see \cite{Santos,JRGW,Deppman1,Lavagno} and references therein). In the meantime the validity  of the nonextensive  $q$-thermodynamics used in such cases was also confirmed \cite{M,M1,M2,M4} and the conditions for its thermodynamical consistency were established \cite{V2,V3,V4,V6}.

In the nonextensive approach one investigates the way in which some selected observables change when one departs from the extensive statistics with value of $q=1$. The goal is to disclose how, and to what extent, these changes are correlated with the possible modifications of the dynamics governing the model considered or with the possible influence of some external factors caused by the surroundings in which formation of dense QCD matter takes place and which is not accounted for in the usual extensive approach. In fact, it is expected that when these factors are gradually identified and their impact is accounted for by a suitable modification of the original model, the value of $|q - 1|$ obtained from comparison with experiment gradually diminishes and $|q-1| = 0$ signals that our improved dynamical model fully reproduces all aspects of the process considered  \cite{MB}.

The investigation of the interplay between these two factors is the subject of our work. However, in the case of the MF models such a procedure is not transparent because of the complexity of the dynamics of MF models (for example, as shown in our nonextensive NJL model \cite{JRGW}, particles acquire dynamical masses which implicitly depend on the nonextensivity parameter).
This prevents a clear interpretation of the role played by the parameter $q$ and its interplay with the dynamics. There is therefore a need to simplify the dynamics, for example by reducing it to a number of well defined (temperature dependent) parameters. Such a possibility is offered by quasi-particle models (QPM) in which the interacting particles (quarks and gluons) are replaced by free quasi-particles. They can be formulated in a number of ways, the most popular approaches are: the model encoding the interaction in the effective masses \cite{QPM3,QPM4}, the model using the Polyakov loop concept \cite{QPM2P,QPM5P} and the model based on the Landau theory of Fermi liquids where the effects of the interaction are modelled by some temperature dependent factors called effective fugacities, $z^{(i)}(T)$, which distort the original Bose-Eistein or Fermi-Dirac distributions \cite{zQPM5,zQPM6,zQPM6a,zQPM11a,zQPM12,zQPM14}.  We will continue to use this model, and call it the $z$-QPM (note that there are also quite a number of other works on the QPM, cf., for example, \cite{QPM-G,QPM-S,QPM-I,QPM-L,QPM-Ba}). This choice is motivated by the fact that in $z$-QPM  the masses of quasi-particles are not modified by the interaction (they do not depend on the fugacities $z^{(i)}$) what allows us to avoid problems encountered in other approaches.

In the $z$-QPM the effective fugacities $z^{(i)}(T)$  ( $z^{(i)} \leq 1$, the $z^{(i)}=1$ correspond to a noninteracting
gas of gluons and quarks) are obtained from fits to lattice QCD results \cite{LQCD1,LQCD3,LQCD4} which serve as a kind of experimental input \cite{zQPM5,zQPM6,zQPM6a,zQPM11a}. Note that the effective fugacities have nothing to do with the usually used fugacities corresponding to the observation of particle number and are therefore not related to the chemical potential; they just encode the effects of the interactions between quarks and gluons.  Because there are problems with allowing for a nonzero chemical potential in lattice simulations \cite{Latt-mu1,Latt-mu3}, the $z$-QPM was initially formulated assuming a vanishing chemical potential, $\mu = 0$. Starting from \cite{zQPM12} a small amount of non-vanishing chemical potential $\mu$ was introduced in the matter sector (to reproduce a realistic equation of state of the QGP), and assumed to be a constant whose value varies between $\mu = 0$ and $100$ MeV (depending on the circumstances, but such that $\mu/T << 1$). However, so far in all fits to lattice data used by the $z$-QPM the chemical potential $\mu$ was neglected.

The aim of this paper (which is an extension of our previous work \cite{q-QPM-RW}) is two-fold. Firstly, after embedding the $z$-QPM in a nonextensive environment characterised by a nonextensive parameter $q$, we investigate the $qz$-QPM created in this way in terms of the changes in the effective fugacities, $z^{(i)} \to z_q^{(i)}$, necessary to fit the same lattice data. Secondly, we use our $qz$-QPM but retain the same effective fugacities $z^{(i)}$ as in the $z$-QMP model and calculate the changes in the densities and pressure induced only by the changes in the nonextensivity $q$. This parallels, in a sense, our nonextensive $q$-NJL model \cite{JRGW} with its dynamics replaced by a phenomenological parametrization in terms of fugacities $z$. However, unlike in the $q$-NJL model, in both cases our investigations are limited to $T$ above the critical temperature $T_c$ because only such are considered in lattice simulations.

Please note that, in terms of dynamics, we do not introduce here any new model. We have just adapted for our purposes the widely known $z$-QPM \cite{zQPM5,zQPM6,zQPM6a,zQPM11a}, accepting its physical motivation which, when combined with its transparency and simplicity, makes this model especially useful for our purposes. However, this also means that the conclusions of this work have, at most, the same level of credibility as those of the $z$-QPM.

The paper is organized as follows. In Section \ref{sec:zQPM} we provide a short reminder of $z$-QPM. Section \ref{sec:qzQPM} contains a formulation of the $qz$-QPM. Our results are presented in  Section \ref{sec:zq} and  Section \ref{sec:Summary} concludes and summarises our work. Technical details are placed in the Appendices \ref{sec:0}, \ref{sec:A}, \ref{sec:B} and \ref{sec:C}.

\section{A short reminder of the $z$-QPM}
\label{sec:zQPM}

We start with a short reminder of the $z$-QPM proposed and used in \cite{zQPM5,zQPM6,zQPM6a,zQPM11a,zQPM12,zQPM14}. It is based on the following effective equilibrium distribution function for quasi-partons ($i=q,~s,~g$ for, respectively, $u$ and $d$ quarks, strange quarks and gluons):
\begin{eqnarray}
n\left[ x^{(i)}\right] &=& \frac{z^{(i)}e\left[-x^{(i)}\right]}{1-\xi\cdot z^{(i)}e\left[-x^{(i)}\right]}
= \frac{1}{\frac{1}{z^{(i)}}e\left[ x^{(i)} \right] - \xi} = \frac{1}{e\left[ \tilde{x}^{(i)}\right] - \xi}, \label{fex-d}\\
x^{(i)} &=& \left\{
 \begin{array}{lll}
  \beta \left[ E_i - \mu^{(i)}\right]~~ &~~ {\rm if}&~~ i=q, s,\\
  \beta E_i ~~&~~ {\rm if} &~~ i= g.
  \end{array}\right.\qquad {\rm and}\qquad \tilde{x}^{(i)} = x^{(i)} - \ln z^{(i)}(\tau)
  \label{x(i)}
\end{eqnarray}
Here $e(x) = \exp(x)$, $\xi = +1$ for bosons and $-1$ for fermions and $\beta = 1/T$. In the $z$-QPM $u$ and $d$ quarks are assumed massless, $E_{i=q}=p$, and strange quarks have mass $m$, $E_s =\sqrt{m^2 + p^2}$; for gluons $E_g = p$. The $z^{(i)} \leq 1$ denote the effective fugacity describing the interactions, they are assumed to depend only on the scaled temperature, $\tau = T/T_{c}$ where $T_{c}$ is the temperature of transition to the deconfined phase of QCD). The dynamics described by the lattice QCD data is encoded in $z^{(i)}$. For  $z^{(i)} = 1$ one has free particles.

The appearance of a chemical potential needs some comment. In the equation of state the fugacity $z$, which is connected with the interactions between particles, changes the pressure $P$ and is therefore connected with the change of the chemical potential $\mu$. It reflects the evolution of the system from some initial state, described by $\mu_0$ and $P_0$, to a state described by $\mu$ and $P$ with $\Delta (\mu) = \mu - \mu_0 = T\ln\left(P/P_0\right)$, which can be derived from the equation of state for constant temperature $T$. For a noninteracting gas where the relative pressure $(P/P_0)\to 1$, this correction vanishes, $\ln(P/P_0)\to 0$. In the $z$-QPM \cite{zQPM5,zQPM6,zQPM6a,zQPM11a,zQPM12,zQPM14} one considers a gas of quarks and gluons above the critical temperature, $T > T_c$, and assumes a quasi-particle description of the lattice QCD equation of state, which in the limit of high temperature ($T\to \infty$) is given by a noninteracting gas of quarks and gluons. The correction $\Delta(\mu)$ is replaced here by the fugacity $z = \exp[ - \Delta(\mu)]$ multiplying distribution function. By analogy to a perfect gas the effective pressure becomes unity in limit of the large $T$ and $z(T\to \infty) \to 1$. Consequently, in the isothermal evolution of a hadron gas for finite temperatures, the chemical potential, or a single particle energy, are corrected by $\Delta(\mu)=T\ln(z)$ . Note that whereas usually the chemical potential $\mu$ enters together with the energy $E$, cf. Eq. (\ref{x(i)}), it can also be associated with the fugacity $x(\tau)$ modifying it by an exponential, temperature dependent, factor:
\begin{equation}
z^{(i)} \to \tilde{z}^{(i)} = z^{(i)}\cdot e\left[ \beta \mu^{(i)}\right]. \label{z-mu}
\end{equation}
The effective fugacity, $\tilde{z}_q^{(i)}$, obtained this way combines the action of the original effective fugacity and that of the chemical potential.

Some remarks concerning the way of the effective fugacities are obtained from the lattice data used in $z$-QPM \cite{zQPM5,zQPM6,zQPM6a,zQPM11a,zQPM12,zQPM14} are in order here. The QCD thermodynamics at high temperature can be described in terms of a grand canonical ensemble which can be expressed in terms of the distribution functions which, in turn, depend on the fugacities, cf. Eq. (\ref{fex-d}). One of the most important quantities calculated on the lattice is pressure. The pressures of the gluons and quarks (expressed as functions of the fugacities) were therefore compared with the corresponding pressures obtained from the lattice data; in this way one gets effective fugacities as functions of scaled temperature, $z(\tau)$ ($\tau = T/T_{cr}$ with $T_{cr}$ being the critical temperature). Because it turns out there is no single universal functional form describing the lattice QCD data over the whole range of $\tau$, the low and high $\tau$ domains were therefore described by different functional forms with the cross-over points at $\tau_g = 1.68$ for gluons and $\tau_q=1.7$ for quarks and were chosen as:
\begin{equation}
z^{(g,q)}(\tau) = a_{(g,q)}\exp\left[ -b_{(g,q)}/\tau^5\right]\cdot\Theta\left( \tau_{(g,q)} - \tau\right) + a'_{(g,q)}\exp\left[ -b'_{(g,q)}/\tau^2\right]\cdot\Theta\left( \tau - \tau_{(g,q)}\right).\label{zzz}
\end{equation}
They were then used to describe the QCD lattice data \cite{LQCD1,LQCD3,LQCD4} with the parameters listed in Table \ref{Table-I}.
\begin{table}[h]
\caption { Numerical values of coefficients $a_{(i)}$, $b_{(i)}$, $a'_{(i)}$ and $b'_{(i)}$ ($i=q,g$) in Eq. (\ref{zzz}) obtained in \cite{zQPM5}.
  }
\vspace*{-0.3cm}
\begin{center}
\begin{tabular}{|c |c| l| l| l| l|}
\cline{1-6}
     $q$ & $(i)$  &~~~ $a_{(i)}$   &~~~ $b_{(i)}$ &~~~  $a'_{(i)}$  &~~~ $b'_{(i)}$   \\
 \hline
$q=1$    &~~$i=g$~~ &~~ $0.803$~~ &~~ $1.837$~~   &~~ $0.978$~~ &~~ $0.942$~~  \\
$q=1$    &~~$i=q$~~ &~~ $0.810$~~   &~~ $1.721$~~   &~~   $0.960$~~ &~~  $0.846$~~  \\ \hline
\end{tabular}
\end{center}
\label{Table-I}
\end{table}

\section{Formulation of the $qz$-QPM}
\label{sec:qzQPM}

To formulate the $qz$-QPM one has to replace the previous extensive effective distribution function for quasi-partons by its nonextensive equivalent,
\begin{equation}
n_q\left[ \tilde{x_q}^{(i)} \right] = \frac{1}{ e_q\left[ \tilde{x_q}^{(i)}\right] - \xi} =
 \frac{e_{2-q}\left[- \tilde{x}_{2-q}^{(i)}\right]}{1 - \xi e_{2-q}\left[- \tilde{x_{2-q}}^{(i)}\right]}\quad {\rm with}\quad
\tilde{x}_q^{(i)} = x^{(i)} - \ln\left[ z_q^{(i)}\right], \label{enqd}
\end{equation}
where
\begin{eqnarray}
e_q(x) = [ 1 + (q-1)x ]^{\frac{1}{q-1}}\stackrel{q \to 1}{\Longrightarrow} e(x)\quad &{\rm and}&\quad
e_{2-q}(-x) = [1 + (1-q)(-x)]^{\frac{1}{1-q}} \stackrel{q \to 1}{\Longrightarrow} e(-x),\label{eqe}\\
e_{q}(-x)\cdot e_{2-q}(x) = 1~ &\stackrel{q \to 1}{\Longrightarrow}&~  e(-x)\cdot e(x) = 1,  \label{qdualq}\\
n_q(x) + n_{2-q}(-x) = - \xi~ &\stackrel{q \to 1}{\Longrightarrow}&~ n(x) + n(-x)= - \xi. \label{nplusnminus}
\end{eqnarray}
Thermodynamical consistency demands that the $n_q(x)$  obtained in this way must be replaced by $n_q(x)^q$ \cite{V4,V6,JR} (this requirement follows from the proper theoretical formulation of the nonextensive thermodynamics provided in \cite{Santos,Deppman1}, cf. Eqs(\ref{qXia}) below).

A comment on the conditions of validity of the $qz$-QPM is in order here. The tacit assumption of the $z$-QPM is that both $x$ and $(-x)$ remain positive, i.e., that $z^{(i)}(\tau) \le 1$ \cite{zQPM5}. However, immersing our system in a nonextensive environment means that some part of the dynamics is now modelled by the parameter $q$, therefore the above constraints are not sufficient because $e_q(x)$ and $e_{2-q}(x)$  must always be nonnegative real valued and the allowed range of $x$ is given by the condition that  $[1 + (q-1)x] \geq 0$ which must be satisfied and which can limit the available phase space \cite{JRGW}. Referring for details to \cite{JR,JRGW} we say only that out of three possibilities of introducing nonextensivity discussed in \cite{JRGW}, only two (one for particles and one for antiparticles) limiting appropriately the available phase space are applicable for our purpose. The third method, which does not limit the available phase space (and which was discussed in detail in \cite{Deppman1}),  introduces some novel dynamical effects, not observed in dense nuclear matter;  therefore we shall not use it here \cite{JRGW}.

Both the form of $n_q(x)$ and the fact that it effectively emerges as $n_q^q(x)$  can be derived from the formulation of the nonextensive thermodynamics in which one starts from the nonextensive partition function $\Xi_q$ (the meaning of the index $i$ and the parameter $\xi$ is the same as in Eqs. (\ref{fex-d}), (\ref{x(i)}) and (\ref{enqd})) taken as \cite{Santos,Deppman1} ($V$ denotes the volume, $i=g,~q,~s$ for, respectively, gluons, light quarks ($u$ and $d$) and strange quarks, and $\nu_i$ are the corresponding degeneracy factors which we take the same as in \cite{zQPM5}: $\nu_g = 16$, $\nu_q = 24$ and $\nu_s=12$):
\begin{equation}
\ln_q \left( \Xi_q\right) = - V \int \frac{d^3 p}{(2\pi)^3}\sum_i \nu_i \xi L_q\left[ \tilde{x}_q^{(i)}\right]\quad
{\rm where}\quad L_q(x) = \ln_{2-q} \left[ 1 - \xi e_{2-q}(-x) \right]. \label{aXi}
\end{equation}
Integrating by parts,
\begin{equation}
\int_0^{\infty}\! p^2 dp\, \ln_{2-q}\left[ 1 - \xi e_{2-q}(-x) \right] =
-\frac{1}{3} \int_0^{\infty}\! p^3dp \frac{\partial}{\partial p} \left\{ \ln_{2-q}\left[ 1 - \xi e_{2-q}(-x) \right]\right\},\nonumber
\end{equation}
and noting that
\begin{equation}
\frac{\partial \ln_{2-q}(x)}{\partial x} = \xi \left[ 1 - \xi e_{2-q}(-x)\right]^{-q}\cdot \left[e_{2-q}(-x)\right]^q = \frac{\xi}{\left[ e_q(x) - \xi\right]^q} = \xi \left[ n_q(x)\right]^q, \nonumber
\end{equation}
one arrives at the following alternative expression for the nonextensive partition function,
\begin{eqnarray}
\ln_q \left( \Xi_q\right) &=& \frac{V}{3} \int \frac{d^3 p}{(2\pi)^3}\sum_i \nu_i p\left[ n_q\left( \tilde{x}_q^{(i)}\right)\right]^q \frac{\partial \tilde{x}_q^{(i)}}{\partial p}, \label{qXia}\\
&&\frac{\partial \tilde{x}_q^{(g,q)}}{\partial p} = \beta, \qquad \frac{\partial \tilde{x}_q^{(s)}}{\partial p} = \beta \sqrt{ 1 + \left(\frac{m}{p}\right)^2 },\nonumber
\end{eqnarray}
with effective distribution functions equal now  $\left[ n_q(x)\right]^q$. A note of caution is necessary here. After closer inspection one realizes that the definition of $e_q(x)$ used in \cite{Santos}, when used together with the duality relation (\ref{qdualq}), leads to $\left[ n_{2-q}\right]^{2-q}$ in Eq. (\ref{qXia}), instead of $n_q^q$ presented in \cite{Santos} (cf., their Eq. (35)). The nonextensive versions of the particle density $\rho_q$ and the energy density, $\varepsilon_q$, are defined, respectively, as (we use Eqs. (\ref{aXi}), (\ref{XEx}), (\ref{pX}) and (\ref{px}) with $\partial x_q/\partial z_q \to \partial x_q/\partial \mu = \beta$),
\begin{eqnarray}
\rho_q &=& \frac{1}{\beta V} \frac{\partial }{\partial \mu} \left[\ln_q\left(\Xi_q\right)\right] = - \frac{1}{\beta }\int \frac{d^3 p}{(2\pi)^3}\sum_i \nu_i \xi \frac{\partial}{\partial \mu}\left\{\ln_{2-q} \left[ 1 - \xi e_{2-q}(-x) \right]\right\} = \nonumber\\
 &=& \int \frac{d^3 p}{(2\pi)^3}\sum_i \nu_i \left[n_q\left(\tilde{x}_q^{(i)}\right)\right]^q = \sum_i \nu_i \rho_q^{(i)}, \label{density}\\
\varepsilon_q &=& - \frac{1}{V}\frac{\partial}{\partial \beta}\left[\ln_q\left(\Xi_q\right)\right] + \sum_i \nu_i \mu^{(i)} \rho_q^{(i)} =\nonumber\\
&=& \int \frac{d^3 p}{(2\pi)^3}\sum_i \nu_i \xi \frac{\partial}{\partial \beta}\left\{\ln_{2-q} \left[ 1 - \xi e_{2-q}(-x) \right]\right\} + \sum_i \nu_i \mu^{(i)} \left[n_q\left(\tilde{x}_q^{(i)}\right)\right]^q =
\nonumber\\
&=& \int \frac{d^3 p}{(2\pi)^3}\sum_i \nu_i \left[ E_i - \frac{\partial}{\partial \beta} \ln z^{(i)}_q(\beta)\right]\cdot \left[n_q\left(\tilde{x}_q^{(i)}\right)\right]^q \label{varepsilon}
\end{eqnarray}
Note that the energy density in our $qz$-QPM depends explicitly on the nonextensivity via the nonextensive particle density and implicitly via a possible $q$-dependence of the effective fugacities mentioned previously. In the extensive limit, $q\to 1$,
Eq. (\ref{varepsilon}) becomes equal to the corresponding equation $(10)$  from the $z$-QPM \cite{zQPM5}.

The physical significance of the effective nonextensive fugacities is best seen when looking at the corresponding nonextensive dispersion relations defined as (cf., Eq. (\ref{varepsilon}))
\begin{equation}
\omega_q^{(i)} = E_i - \frac{\partial}{\partial \beta} \ln\left[ z_q^{(i)}(\beta)\right] = E_i + T^2\left[ \frac{1}{z_q^{(i)}} \frac{\partial z_q^{(i)}}{\partial T}\right]. \label{DR}
\end{equation}
Note that the masses of the quasiparticles remain intact and the single quasiparticle energies $\omega_q^{(i)}$ are modified only by the action of the effective fugacities, $z_q^{(i)}(T)$. In both extensive and nonextensive cases this results in some additional contributions to the quasiparticle energies which can be interpreted as coming from the collective excitations. They occur because of the temperature dependence of the effective fugacities (deduced from the lattice calculations) which can be interpreted as representing the action of the gap equation in \cite{JRGW} taken at constant energy $E_i$.

Because in the version of $q$-thermodynamics used here all thermodynamic relations are preserved, the pressure $P_q$ is given by the  usual thermodynamic relation,
\begin{equation}
P_q\beta V = \ln_q\left(\Xi_q\right). \label{P}
\end{equation}
From Eqs. (\ref{varepsilon}) and (\ref{P}) one gets an expression for the trace anomaly ($\beta = 1/T$)
\begin{equation}
\mathbb{T}_q = \frac{\varepsilon_q - 3 P_q}{T^4} = T\frac{\partial}{\partial T}\left( \frac{P_q}{T^4}\right) = -\beta\frac{\partial }{\partial \beta}\left( P_q\beta^4\right). \label{TraceA}
\end{equation}

\section{Results}
\label{sec:zq}

We shall now calculate the nonextensive effective fugacities, $z_q(\tau)$, which, for a given value of the nonextensivity parameter $q$, reproduce the original $z$-QPM results \cite{zQPM5}. These results, in turn, were obtained from a comparison with the lattice QCD simulations from \cite{LQCD1} using relation (\ref{P}) to match the pressures in the $z$-QPM and in the lattice QCD simulations. Note that this procedure assumes in fact that the trace anomaly in the $z$-QPM (Eq. (\ref{TraceA}) with $q=1$) is the same as that resulting from the QCD lattice data \cite{zQPM5}. We adopt the same procedure and use Eq. (\ref{P}) to match the pressures calculated, respectively, for $q=1$ (as in in \cite{zQPM5}) and for $q \neq 1$,
\begin{figure}[h]
\resizebox{0.5\textwidth}{!}{%
  \includegraphics{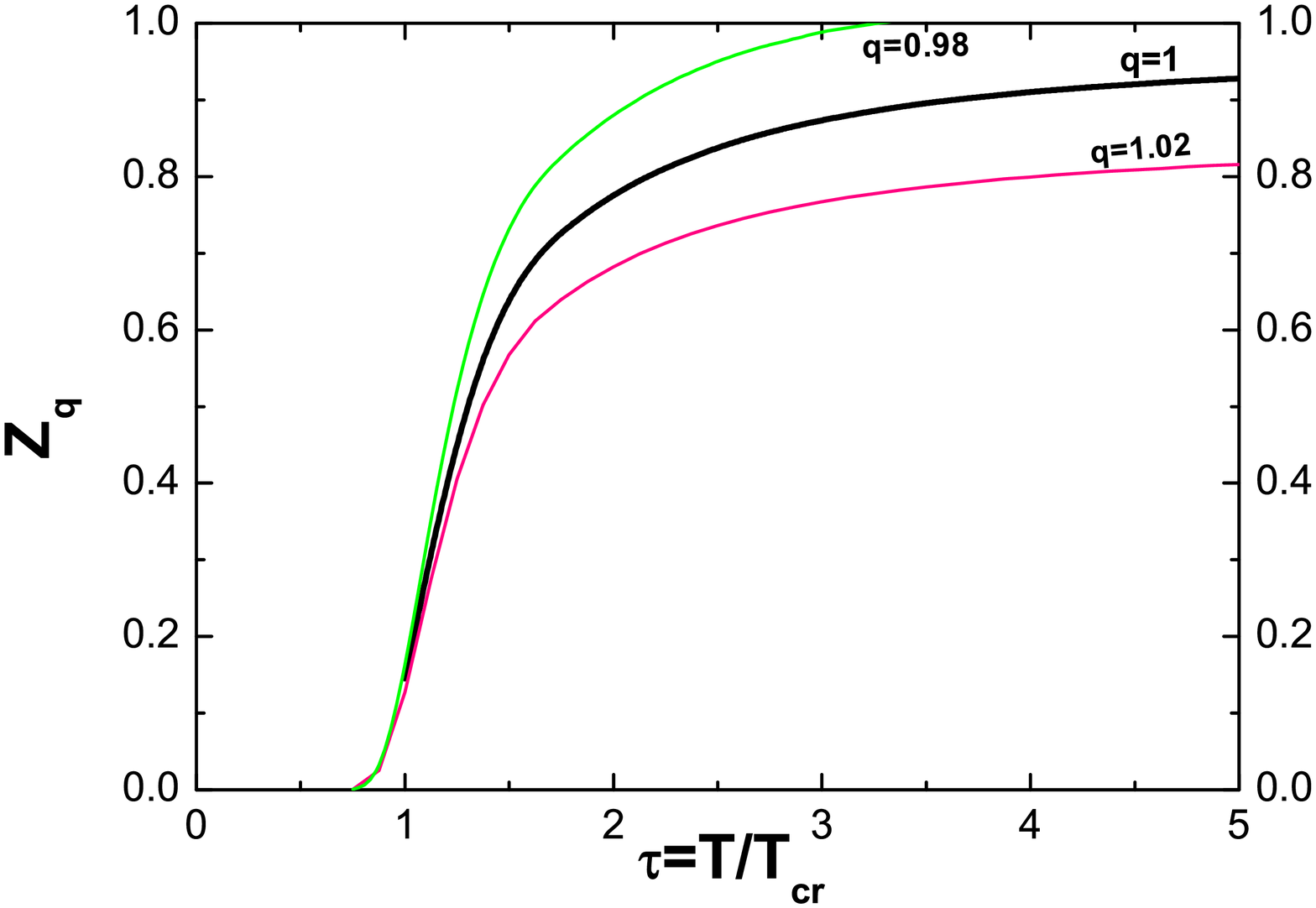}
}
\resizebox{0.51\textwidth}{!}{%
  \includegraphics{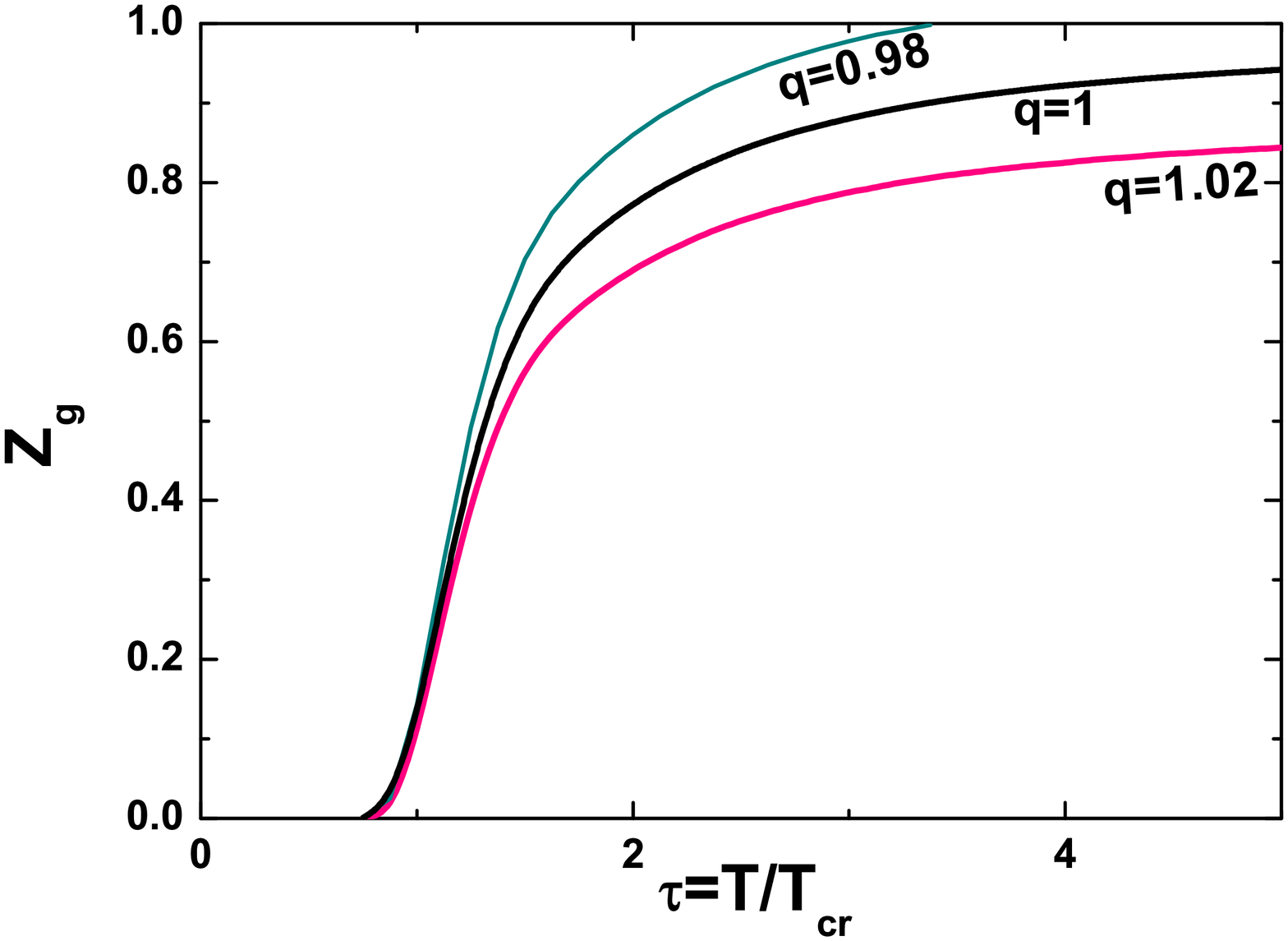}
}\\
\resizebox{0.5\textwidth}{!}{%
  \includegraphics{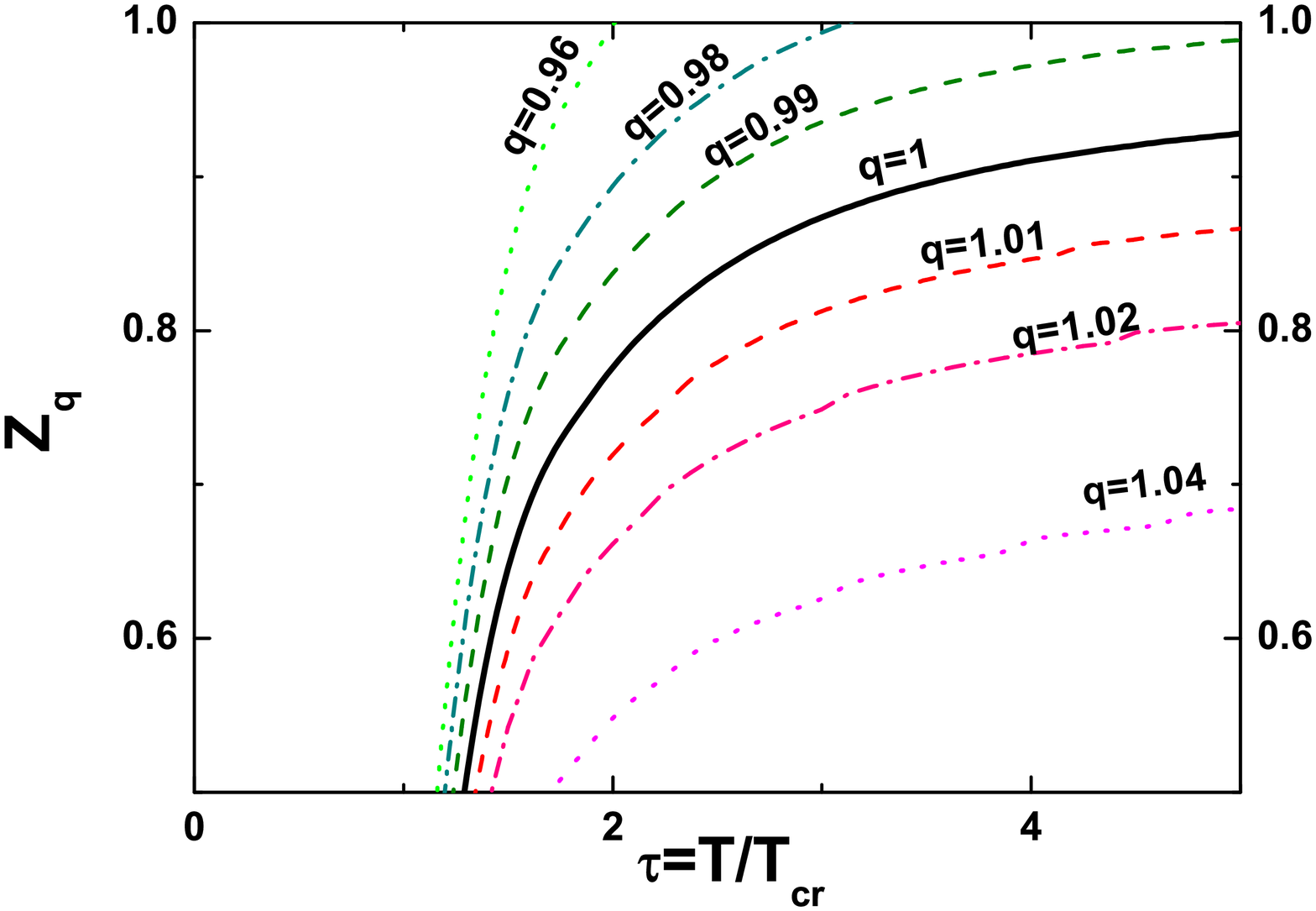}
}
\resizebox{0.51\textwidth}{!}{%
  \includegraphics{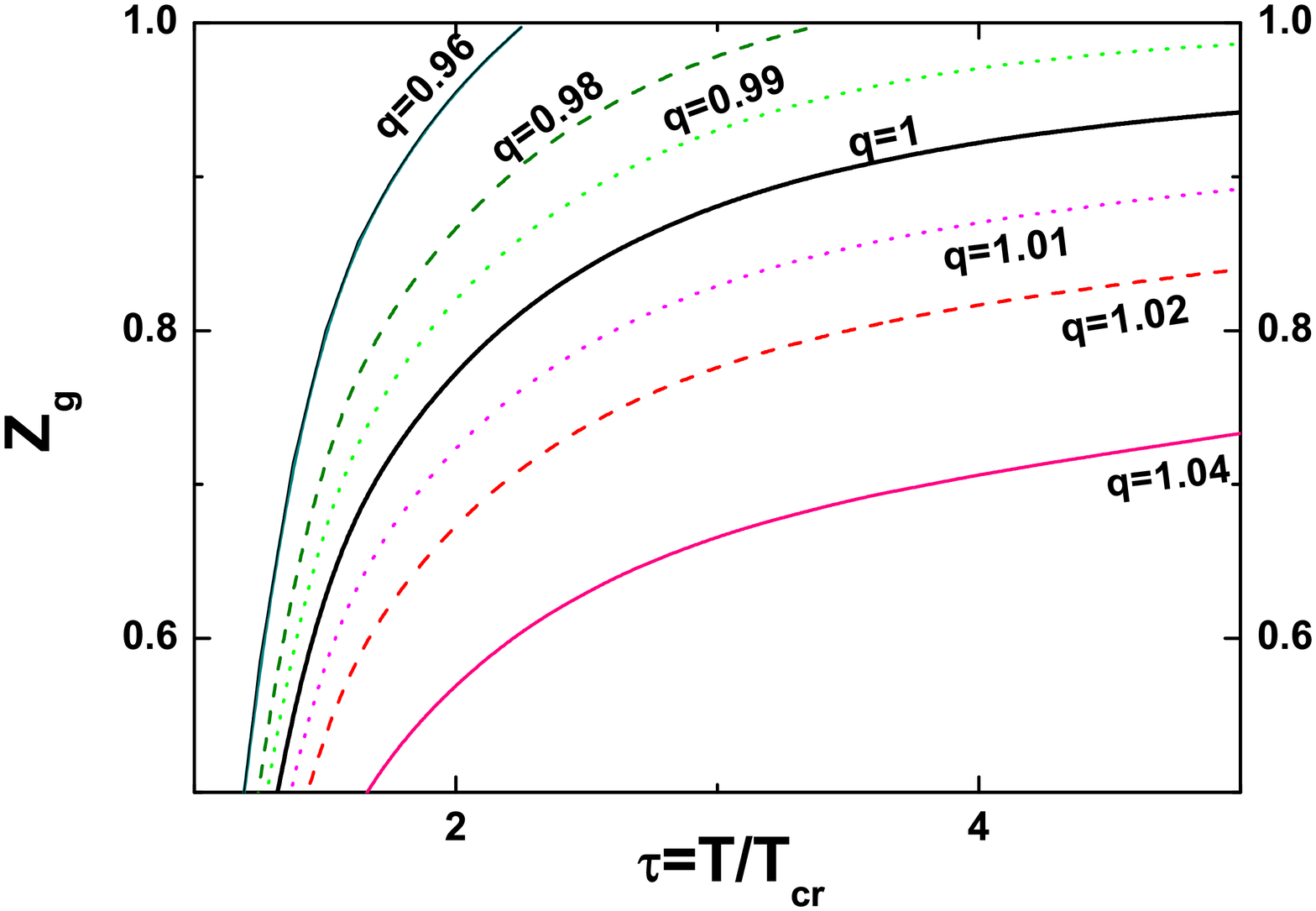}
}
\caption{(Color online) Upper panels: Results for $z_q^{(q)}(\tau)$ and $z_q^{(g)}(\tau)$ as a function of the scaled temperature $\tau = T/T_{c}$ (calculated for $\mu=0$). Lower panels: As above but shown in more detail and with an enlarged range of the nonextensivity parameter $q$.}
\label{zqg-1}
\end{figure}
\begin{equation}
P_{q=1}(T) = P_q(T) \label{P_Pq}
\end{equation}
(it is tacitly assumed that in both extensive and nonextensive environments the temperature $T$ remains the same). This means that in our case the trace anomaly remains the same as in the $z$-QPM (and as in the lattice data) and does not depend on the nonextensivity. To this end the following conditions must be satisfied:
\begin{equation}
\int_0^{\infty} dp p^2 \ln \left[ 1 - e\left( -\tilde{x}^{(g)} \right)\right] = \int_0^{\infty}\,\, dp p^2 \ln_{2-q}\left[ 1 - e_{2-q}\left( -\tilde{x}_q^{(g)}\right)\right] \Theta\left(p;g)\right),\label{gluons}
\end{equation}
for gluons (with $\nu_g = 16$) and
\begin{eqnarray}
\hspace{-10mm}&& \nu_q \int_0^{\infty} dp p^2 \ln \left[ 1 + e\left( -\tilde{x}^{(q)} \right)\right] + \nu_s \int_0^{\infty} dp p^2 \ln \left[ 1 + e\left( -\tilde{x}^{(s)} \right)\right] = \nonumber\\
\hspace{-10mm}&& = \nu_q \int_0^{\infty} dp p^2 \ln_{2-q}\left[ 1 + e_{2-q}\left( -\tilde{x}_q^{(q)}\right)\right]\Theta(p;q) +
\nu_s \int_0^{\infty} dp p^2 \ln_{2-q}\left[ 1 + e_{2-q}\left( -\tilde{x}_q^{(s)}\right)\right]\Theta(p;s), \label{quarks}
\end{eqnarray}
for quarks. They give us the $\tau$ and $q$-dependent relations between the extensive fugacities obtained in \cite{zQPM5}), $z^{(i)}(\tau)$ (which are our input), and the nonextensive fugacities, $z_q^{(i)}(\tau)$ (which are our results). The function $\Theta(q,q)$ defines the allowed phase space; its details are presented in Appendix \ref{sec:0}.

\begin{table}[h]
\caption { Numerical values of the coefficients $a_{(i)}$, $b_{(i)}$, $a'_{(i)}$ and $b'_{(i)}$ ($i=q,g$) in Eq. (\ref{zzz}) when used for different values of $q$ resulting in the curves displayed in the lower panels of Fig. \ref{zqg-1}.}
\vspace*{-0.3cm}
\begin{center}
\begin{tabular}{|c |c| l| l| l| l|}
\cline{1-6}
     $q$ & $(i)$  &~~~ $a_{(i)}$   &~~~ $b_{(i)}$ &~~~  $a'_{(i)}$  &~~~ $b'_{(i)}$   \\
 \hline
 \hline
$q=0.96$ &~$i=g$~~ &~~ $0.985$~~ &~~ $1.581$~~  &~~ $1.168$~~  &~~ $0.860$~~    \\ 
$q=0.96$ &~$i=q$~~ &~~ $1.030$~~ &~~ $1.510$~~  &~~ $1.200$~~  &~~ $0.747$~~    \\ \hline
\hline
$q=0.98$ &~$i=g$~~ &~~ $0.897$~~ &~~ $1.702$~~  &~~ $1.078$~~  &~~ $0.870$~~    \\ 
$q=0.98$ &~$i=q$~~ &~~ $0.924$~~ &~~ $1.603$~~  &~~ $1.073$~~  &~~ $0.770$~~    \\ \hline
\hline
$q=0.99$ &~$i=g$~~ &~~ $0.850$~~ &~~ $1.760$~~  &~~ $1.028$~~  &~~ $0.904$~~    \\ 
$q=0.99$ &~$i=q$~~ &~~ $0.867$~~ &~~ $1.662$~~  &~~ $1.018$~~  &~~ $0.799$~~    \\ \hline
\hline
$q=1.01$ &~$i=g$~~ &~~ $0.753$~~ &~~ $1.916$~~  &~~ $0.927$~~  &~~ $0.990$~~    \\ 
$q=1.01$ &~$i=q$~~ &~~ $0.751$~~ &~~ $1.791$~~  &~~ $0.896$~~  &~~ $0.879$~~    \\ \hline
\hline
$q=1.02$ &~$i=g$~~ &~~ $0.704$~~ &~~ $2.006$~~  &~~ $0.876$~~  &~~ $1.059$~~    \\ 
$q=1.02$ &~$i=q$~~ &~~ $0.694$~~ &~~ $1.862$~~  &~~ $0.835$~~  &~~ $0.925$~~    \\ \hline
\hline
$q=1.04$ &~$i=g$~~ &~~ $0.600$~~ &~~ $2.221$~~  &~~ $0.766$~~  &~~ $1.180$~~    \\ 
$q=1.04$ &~$i=q$~~ &~~ $0.580$~~ &~~ $2.061$~~  &~~ $0.712$~~  &~~ $1.050$~~    \\ \hline
\end{tabular}
\end{center}
\end{table}

Figs. \ref{zqg-1} shows the resulting effective fugacities, $z_q = z_q^{(q)}(\tau)$ and $z_g=z_q^{(g)}(\tau)$, as functions of the scaled temperature, $\tau = T/T_{c}$. They can be fitted using the same parametrization as before, i.e., Eq. (\ref{zzz}), with the parameters displayed in Table II. Since the values of $z_{q=1}^{(i)}$ obtained in \cite{zQPM5} were obtained assuming $\mu = 0$, the same assumption was used in obtaining our $z_q^{(i)}$ here. Note that for the nonextensivites $q$ used here the changes in the fugacities are small,
\begin{equation}
\delta z_q^{(q,g)} = z_q^{(q,g)} - z_{q=1}^{(q,g)} < 1, \label{deltaz}
\end{equation}
and can be approximated (with very good accuracy of a few percent) by (cf. Appendix {\ref{sec:A}),
\begin{equation}
\delta z_q \simeq z_{q=1} (1-q) \cdot F\left(q=1,z_{q=1}\right)\quad {\rm where}\quad  F = \frac{\int_0^{\infty} dp p^2 \left\{ \ln^2[ 1 - \xi e(-x;z)] + n(x;z) x^2\right\}}{2\int_0^{\infty} dp p^2 n(x;z)}. \label{Approximation}
\end{equation}
Eq. (\ref{Approximation}), together with Fig. \ref{zqg-1}, allows for a better understanding of  interrelation between the dynamics (represented by the fugacities $z$) and the nonextensivity  described by $q$. The central point is that all the $z_q$ must describe the lattice QCD data (directly for $q=1$ in $z$-QPM and indirectly for $q\neq 1$ in $qz$-QPM, where they are forced to reproduce the results of $z$-QPM). The $\tau$-dependence of $z(\tau)$ starts from small values (corresponding to strong attraction) towards $z=1$ (corresponding to free, noninteracting particles). The case of $z > 1$ would formally mean the emergence of repulsive forces and is not allowed in $z$-QPM, therefore we shall also keep this limitation  in our $qz$-QPM. The replacement of extensive media by not extensive means adding some repulsive interaction (in the case of $q<1$) or an attractive one (for $q>1$). Therefore, in the first case it must be compensated by an increase in $z$ (i.e., $\delta z_q > 0$) and in the second case by a decrease (i.e., $\delta z_q <0$). Note now that whereas in the latter case we have $z_q(\tau) < z_{q=1}(\tau) < 1$, in the former there is limiting value of $\tau=\tau_{lim}(q)$, depending on $q$, for which $z_q\left( \tau_{lim}\right) = 1$. This means that for $\tau > \tau_{lim}(q)$ the attraction represented by $z(\tau)$ is already too weak to compensate the repulsion introduced by $q<1$. The value of $\tau_{lim}$ diminishes with the increase of this repulsion (i.e., with the increase of $|q-1|$). Not wanting to introduce the problem of repulsion we limit our considerations to $\tau > \tau_{lim}$ only.

\begin{figure}[h]
\resizebox{0.5\textwidth}{!}{%
  \includegraphics{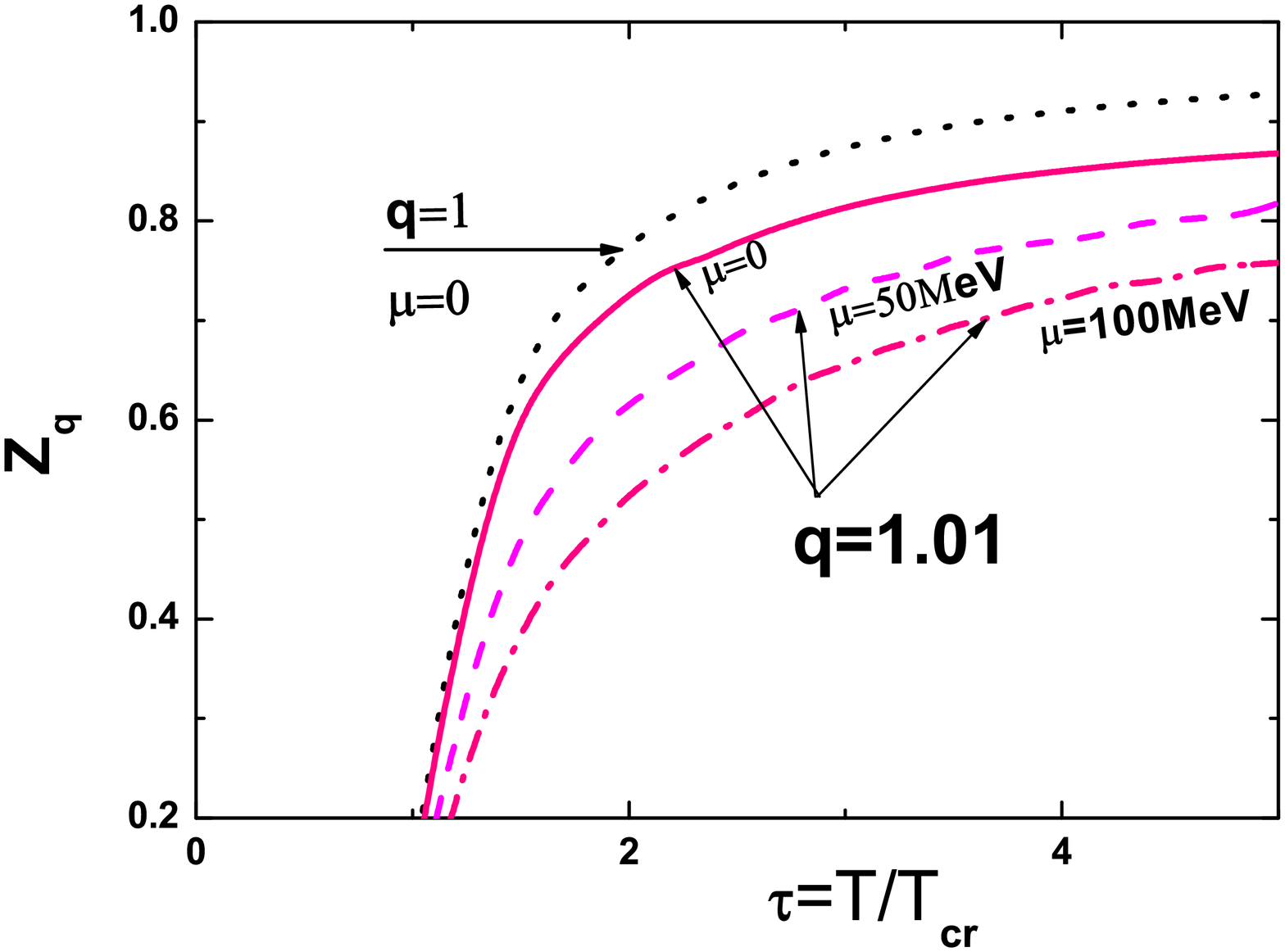}
}
\resizebox{0.51\textwidth}{!}{%
  \includegraphics{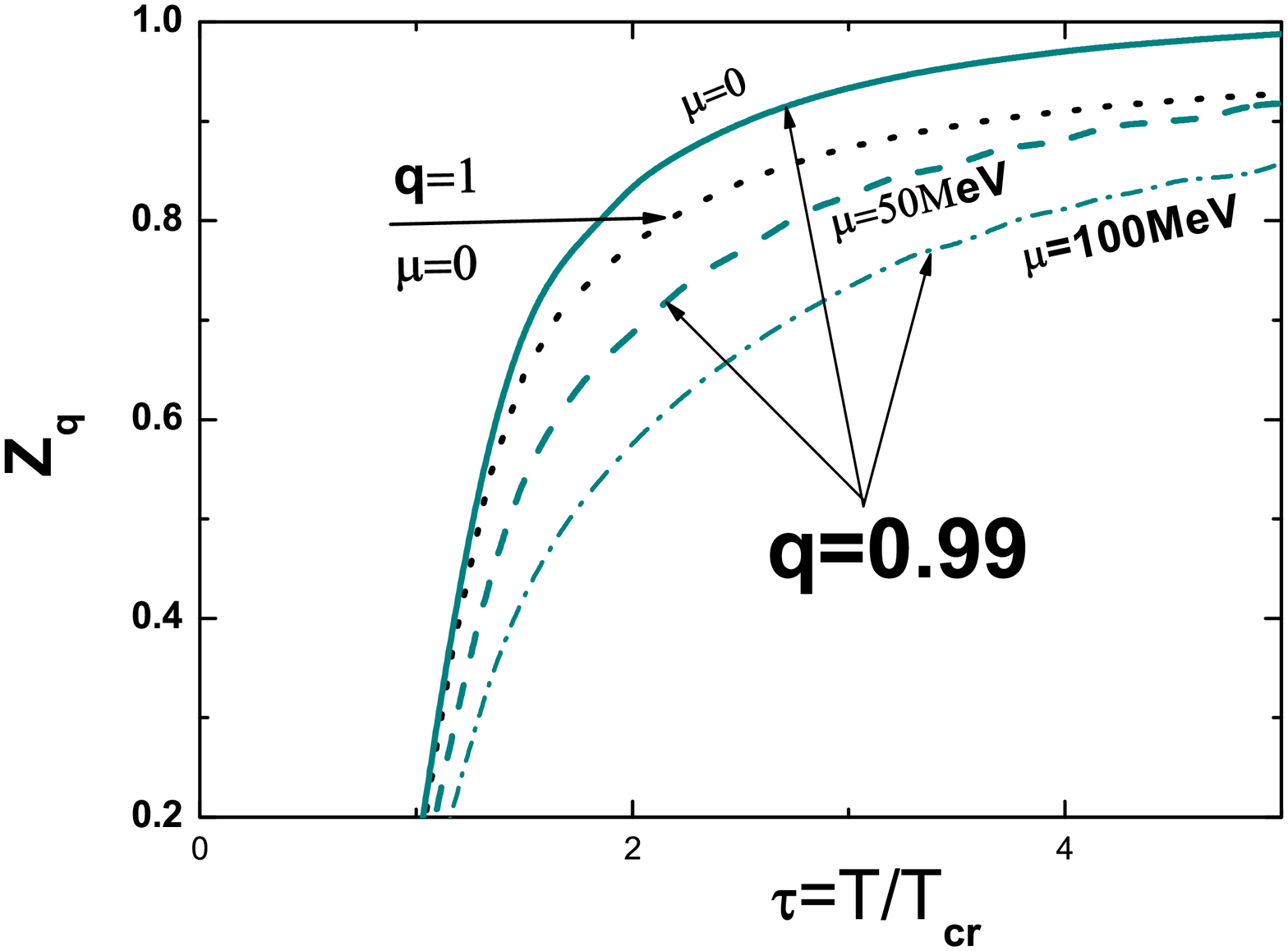}
}
\caption{(Color online) Illustration of the changes introduced by the chemical potential $\mu$ for $q = 1.01$ and $q=0.99$.}
\label{zqmu}
\end{figure}
So far, results for $z(\tau)$ and $z_q(\tau)$ have been obtained with $\mu = 0$. The formal introduction of the chemical potential $\mu$ in $z$-QPM \cite{zQPM12} makes $z$-QPM more flexible and applicable to possible future lattice QCD data with the chemical potential accounted for. Following this new development in $z$-QPM we have also formally introduced $\mu$ into our $z$-QPM. We can therefore check what would be the value of our $z_q$ in the case when part of the dynamics is shifted from fugacity $z$ to the chemical potential $\mu$. Eq. (\ref{z-mu}) shows the  effective fugacity with the chemical potential included. It is visualized in Fig. \ref{zqmu} where we plot a number of results for different values of the chemical potential $\mu$ and for two values of the nonextensivity parameter: $q = 0.99$ and $q=1.01$. As one can see, nonzero $\mu$ diminishes the real values of the fugacity because, according to Eq. (\ref{z-mu}) (valid also in a nonextensive environment with $\tilde{z} \to \tilde{z}_q$), the effective value $\tilde{z}_q$ now contains an  exponential factor greater than unity, which modifies the original fugacity $z$.

\begin{figure}[t]
\resizebox{0.5\textwidth}{!}{%
  \includegraphics{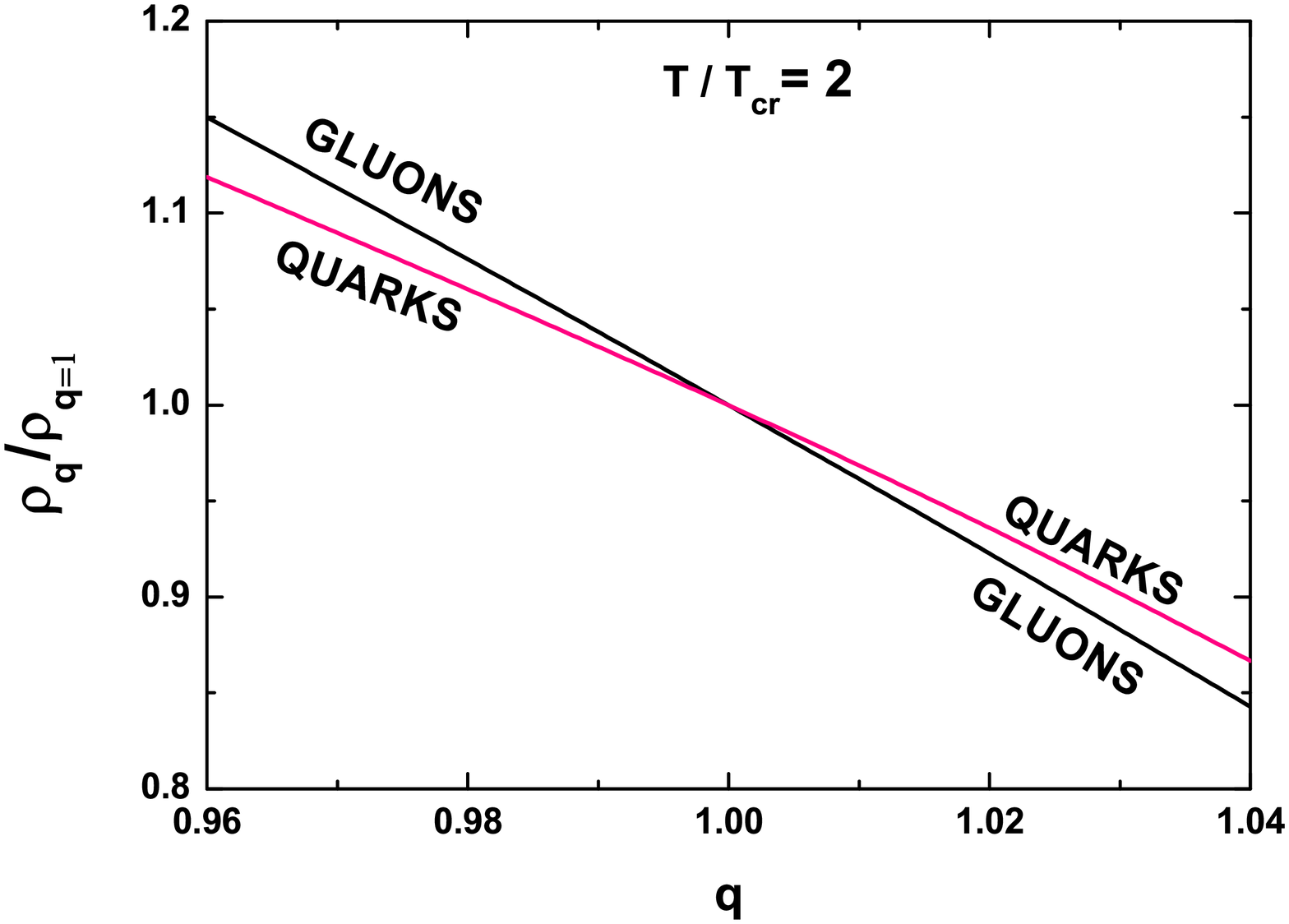}
}
\resizebox{0.5\textwidth}{!}{%
  \includegraphics{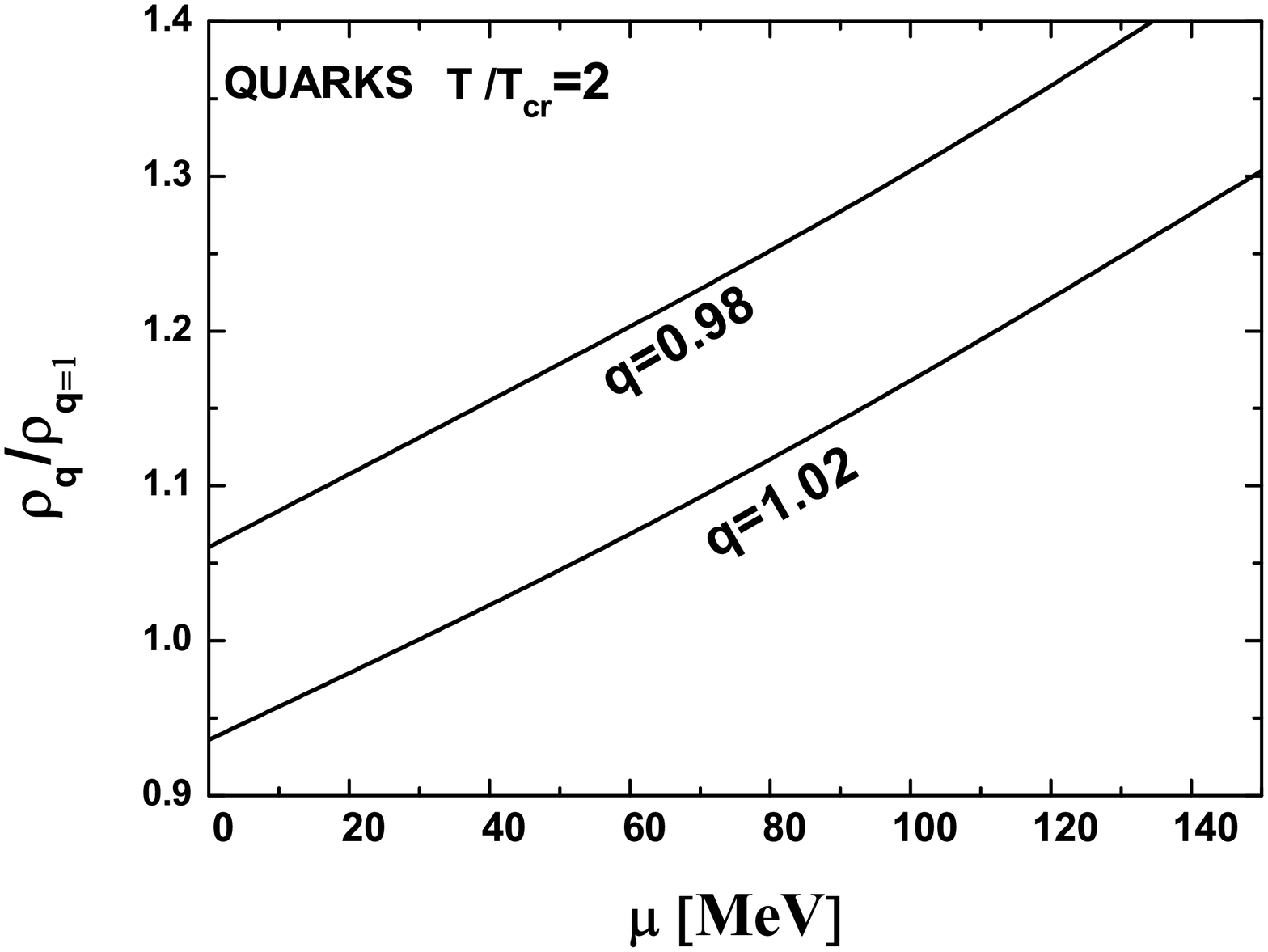}
}
\caption{(Color online) Left panel: Relative density, $\rho_q/\rho_{q=1}$, of quarks and gluons as a function of the nonextensivity $q$ for $\mu =0$. Right panel: Dependence of the relative density, $\rho_q/\rho_{q=1}$, on the chemical potential $\mu$.}
\label{mu-density}
\end{figure}
\begin{figure}[h]
\resizebox{0.5\textwidth}{!}{%
  \includegraphics{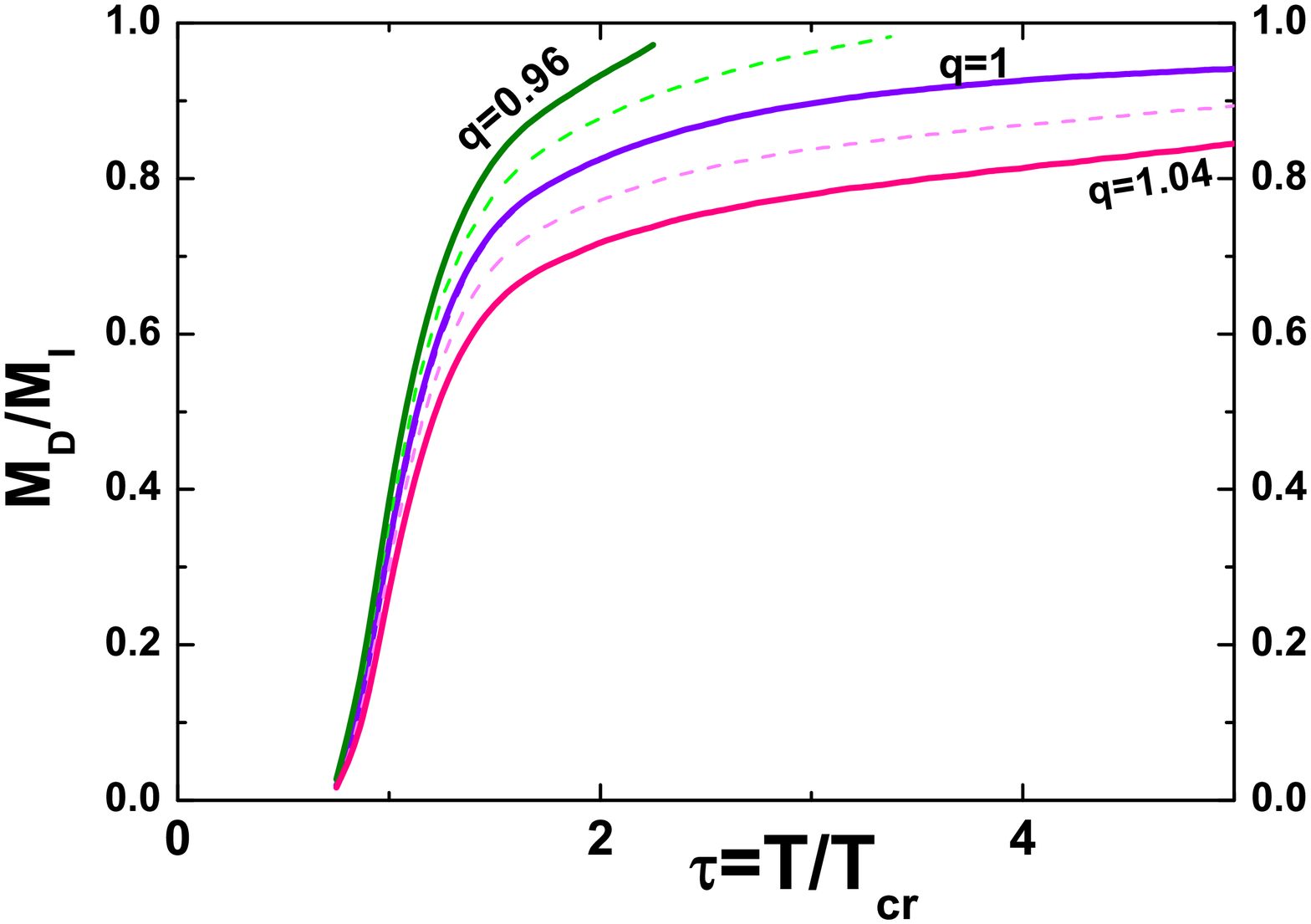}
}
\resizebox{0.51\textwidth}{!}{%
  \includegraphics{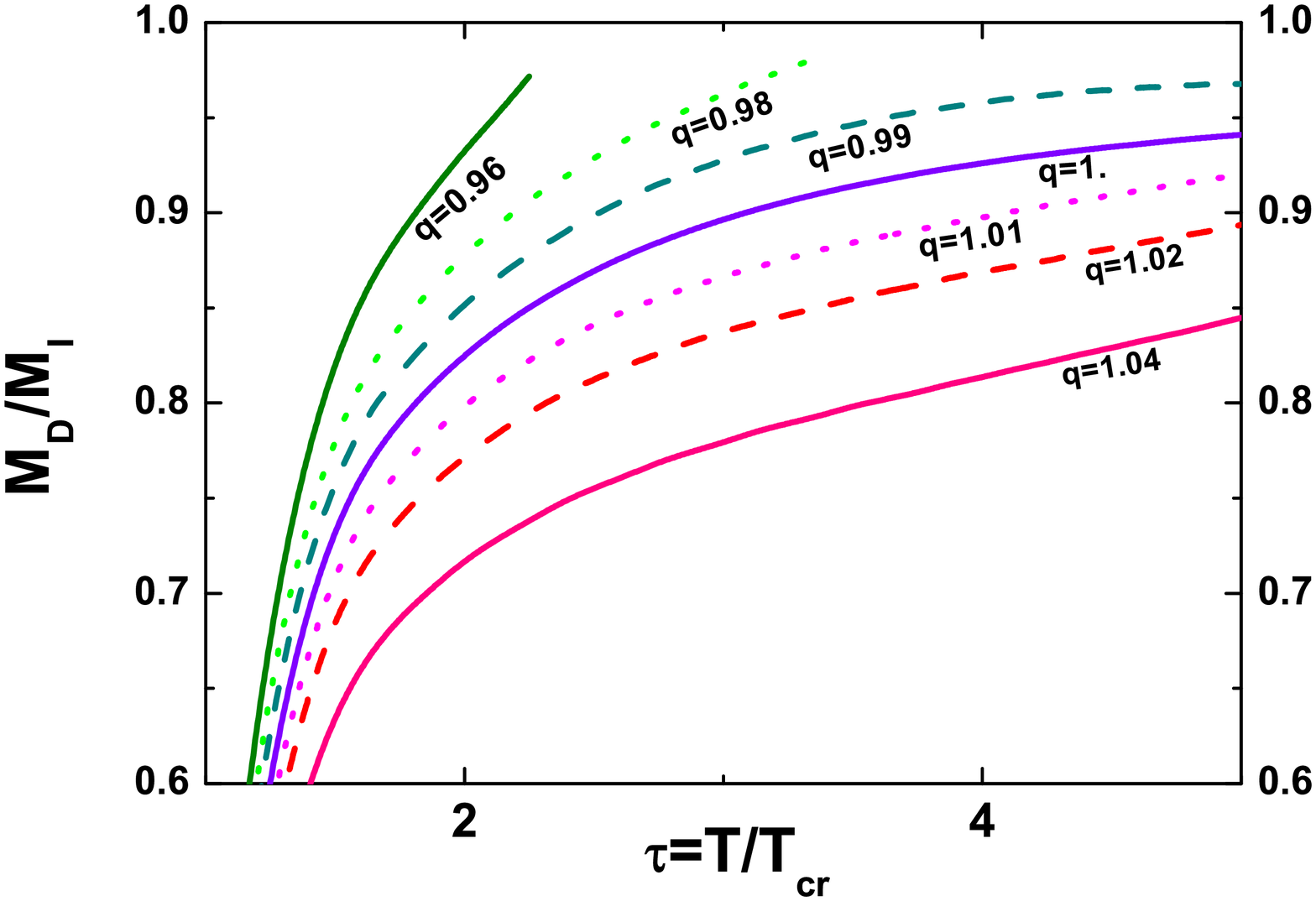}
}
\caption{(Color online) Left panel: Results for the ratio $M_D/M_D^I$ of the Debye masses (as defined by Eqs. (\ref{MD}) and (\ref{MDI})), respectively) in the nonextensive environment as functions of the scaled temperature, $\tau = T/T_{c}$ for $q=0.96,~1,~1.04$, calculated for $\mu=0$. Right panel: The same as above but shown in more detail and with an enlarged range of the nonextensivity parameter $q$.}
\label{f-MD}
\end{figure}
The introduction of the chemical potential $\mu$  also changes the $q$-dependence of the relative density of the quarks, $R_{\rho} = \frac{\rho_q}{\rho_{q=1}}$, where $\rho_q$ is given be Eq. (\ref{density}). As can be seen in the left panel of Fig. \ref{mu-density}, in a nonextensive environment one observes a clear separation of the situations with $R_{\rho} > 1$ and $R_{\rho} <1$. The first occurs for $q<1$ and  the observed increase of density is consistent with lowering of the entropy which, in turn, is connected with the tighter packing of the quarks in this case \cite{JRGW}. The second occurs for $q > 1$ and the picture is reversed; it is consistent with an increase of the entropy and with looser packing of the quarks in this case. Note that this behaviour of $R_{\rho} = R_{\rho}(q)$ is fully consistent with the behaviour of the nonextensive fugacities presented in Fig. \ref{zqg-1}. Essentially the same result can be obtained using the linear approximation of the $n_q^q$ in $(q-1)$ as given by Eqs. (\ref{eq-qnq}) and (\ref{Deq-qnq}). Using now the same values of $z_q$ but adding some amount of the chemical potential $\mu$ yields the results shown in the right panel of Fig. \ref{mu-density}.  We observe some increase of the relative density with $\mu$ with a possible trace of a small upper bending.

We shall now calculate the modifications of partonic charges in a hot QCD medium embedded in a nonextensive environment calculating the corresponding Debye mass, $M_{qD}$. Following \cite{zQPM5} we use for the extensive Debye mass the expression derived in semiclassical transport theory in which $M_D$ is given in terms of equilibrium parton distribution functions ($N_c$ denotes the number of colors):
\begin{equation}
M^2_D = - 2N_cQ^2\int\frac{d^3p}{8\pi^3}\partial_p n^{(g)} - Q^2\int\frac{d^3p}{8\pi^3}\partial_p\left( 4 n^{(q)}+2 n^{(s)}\right) = \frac{N_c Q^2}{\pi^2} n^{(g)} - Q^2 \left( 2 n^{(q)}
 +  n^{(s)}\right). \label{MD}
\end{equation}
In the nonextensive environment described by the nonextensivity parameter $q$ we simply replace $n^{(i=g,q,s)}(x;z)$ by $\left[ n_q^{(i)}\left( x;z_q\right)\right]^q$. In Fig. \ref{f-MD}, following \cite{zQPM5}, we present the ratio of $M_D/M_D^I$ where $M_D^I$ denotes the Debye mass for the ideal EOS case (i.e., with $z_g=1$ and $z_q=1$) which, following \cite{zQPM5}, equals
\begin{equation}
M_D^I = Q T \sqrt{ \frac{N_c}{3} + \frac{1}{2} - \frac{m^2}{4\pi^2 T^2} \ln 2}. \label{MDI}
\end{equation}
Note that because the Debye mass  is essentially a combination of the densities of quarks and gluons the above results resemble those for the effective fugacities and all previous remarks also apply here.

We now proceed to the second part of our work in which we keep the original dynamics of the $qz$-QPM intact  using the same effective fugacities as in \cite{zQPM5} (i.e., we assume that $z_q{(i)} \to z^{(i)}$ as given by Eq. (\ref{zzz}) with the parameters listed in Table I). This parallels to some extent our approach in the nonextensive $q$-NJL model \cite{JRGW} (but now the dynamics is simplified and represented by the temperature dependent fugacities $z(\tau)$ reproducing the lattice QCD results) and allows us to investigate the sensitivity of the selected observables to the nonextensive environment only. The only drawback is the limitations in the temperatures allowed because the fugacities are only defined for $\tau > 1$, i.e., above the critical temperature $T_c$.

\begin{figure}[h]
\resizebox{0.5\textwidth}{!}{%
  \includegraphics{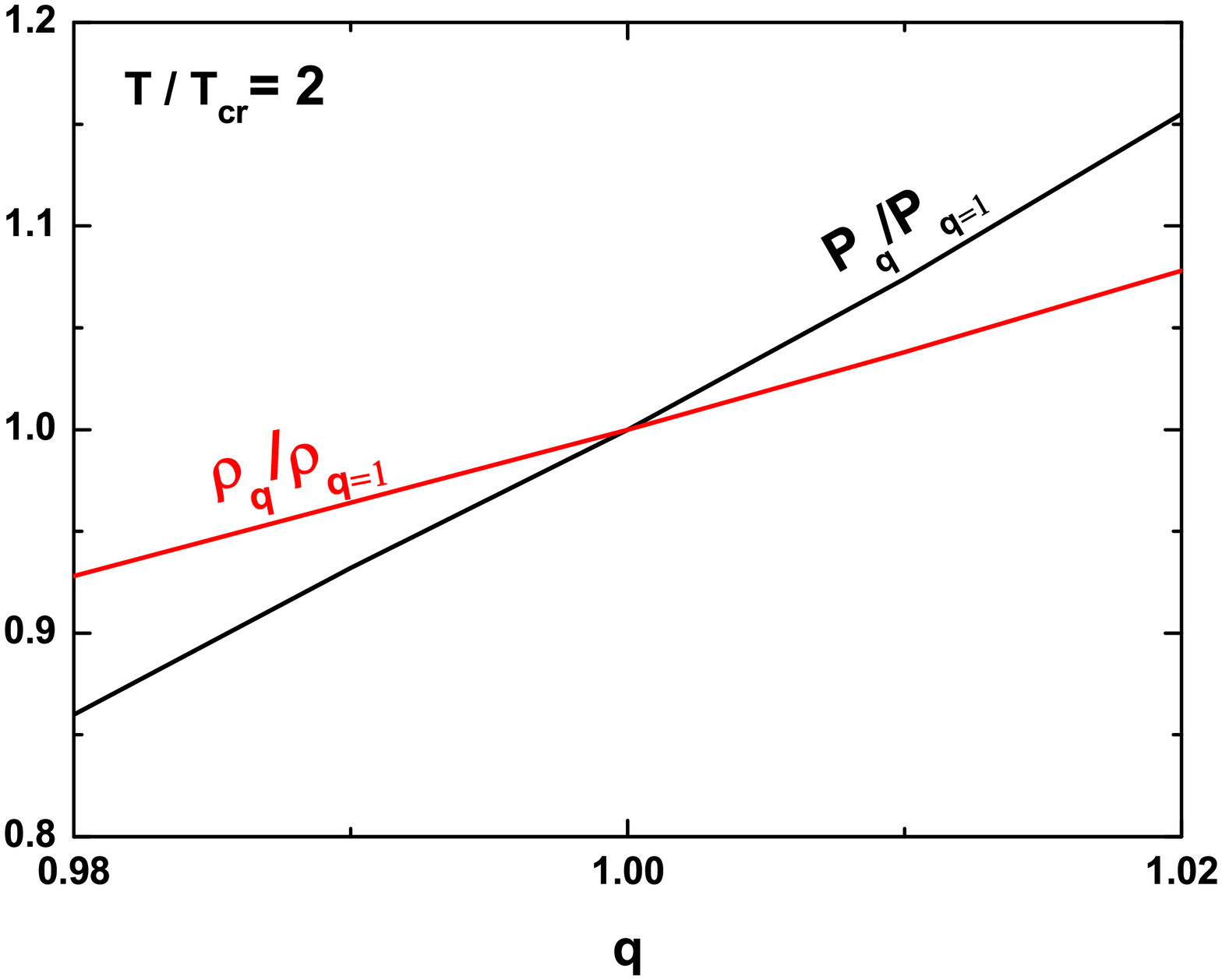}
}
\resizebox{0.51\textwidth}{!}{%
  \includegraphics{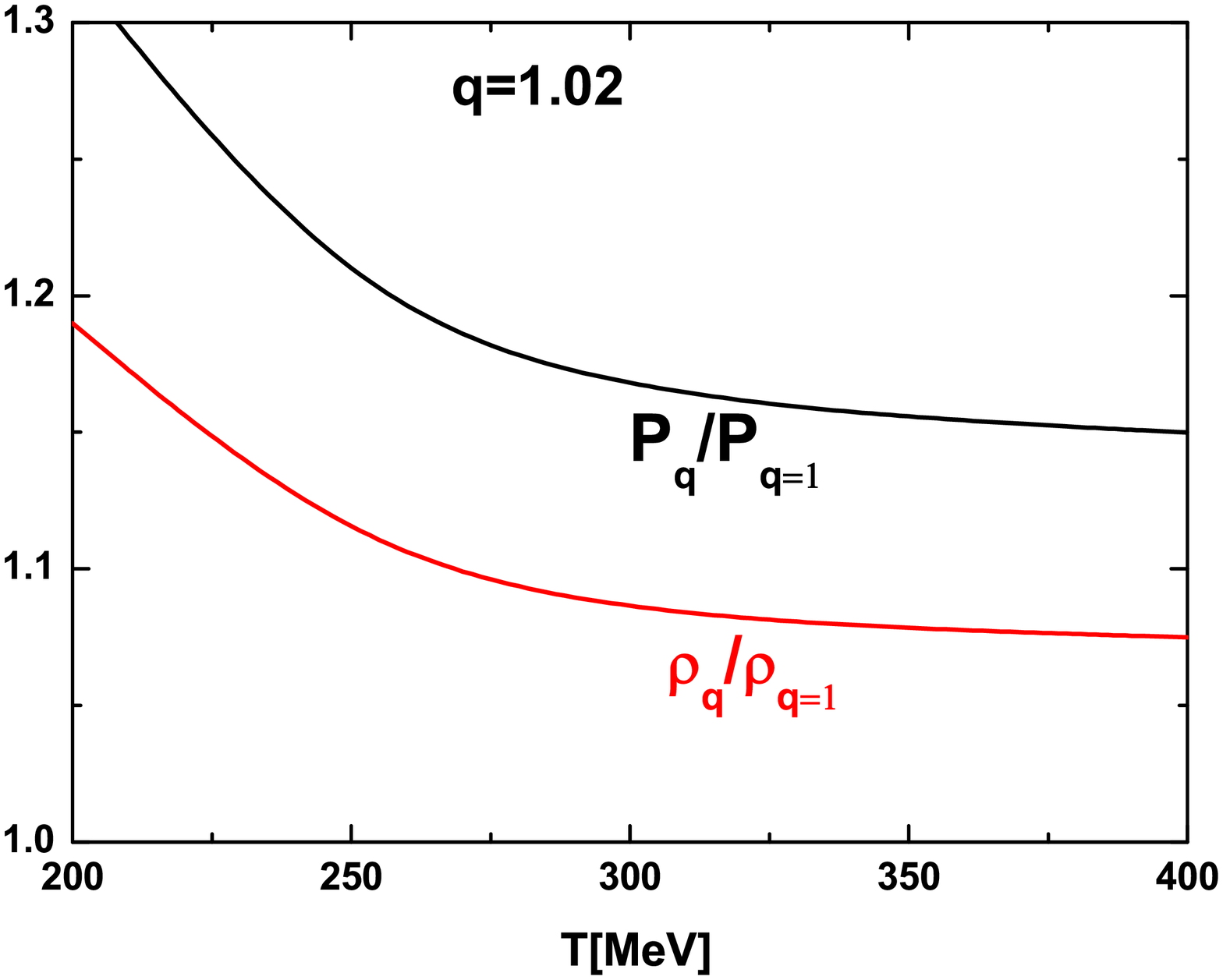}
}
\caption{(Color online) Dependencies of relative pressure $P_q/P_{q=1}$ and density $\rho_q/\rho_{q=1}$ on the nonextensivity parameter $q$ at fixed temperature (left panel) and on the temperature $T$ for fixed $q$ (right panel).}
\label{Original}
\end{figure}

To start with we present in Fig. \ref{Original} the corresponding relative pressure $P_q/P_{q=1}$ and relative density $\rho_q/\rho_{q=1}$ as functions of the nonextensivity parameter $q$ at fixed temperature (left panel) and their dependencies on $T$ for some fixed nonextensivity $q$ (right panel). Note that whereas before the pressure was assumed to be the same for extensive and nonextensive environments, $P_q/P_{q=1} = 1$, it now increases linearly with $q$ in the same way as in the $q$-NJL model \cite{JRGW}. The relative density also increases with the nonextensivity $q$, contrary to its previous behaviour  (demonstrated in the left panel of Fig. \ref{mu-density}) where it decreased with $q$.

We now proceed to the trace anomaly, $\mathbb{T}_q$,  Eq. (\ref{TraceA}). Note that now it acquires some explicit dependence on the nonextensive parameter $q$. Using the definitions of energy density, $\varepsilon_q$, Eq. (\ref{varepsilon}) and pressure $P_q$ (Eq. (\ref{P}) (and additionally Eqs. (\ref{qXia}) and (\ref{x(i)})), we obtain that
\begin{eqnarray}
\mathbb{T}_q &=& \beta^4 \int \frac{d^3 p}{(2\pi)^3}\sum_i \nu_i \left[ E_i - p \frac{\partial \tilde{x}^{(i)}}{\partial p} - \frac{\partial}{\partial \beta} \ln z^{(i)}(\beta)\right]\cdot \left[n_q\left(\tilde{x}^{(i)}\right)\right]^q =\nonumber\\
&=& - \beta^4 \frac{\partial}{\partial \beta} \ln z^{(i)}(\beta)\cdot \int \frac{d^3 p}{(2\pi)^3}\sum_i \nu_i \left[n_q\left(\tilde{x}^{(i)}\right)\right]^q. \label{tranomalyq}
\end{eqnarray}
The change in the trace anomaly generated by the nonextensivity $q$ is given by (see Eqs. (\ref{eq-qnq}) and (\ref{Deq-qnq}))
\begin{eqnarray}
\Delta\left[ \mathbb{T}_q\right] &=& \mathbb{T}_q - \mathbb{T}_{q=1} =
- \beta^4 \frac{\partial}{\partial \beta} \ln z^{(i)}(\beta)\cdot \int \frac{d^3 p}{(2\pi)^3}\sum_i \nu_i \Delta^{(i)}\left[ n^q_q;n\right]; \label{TA-1}\\
\!\!\!\!\!\Delta^{(i)}\left[ n^q_q;n\right]\!\! &=&\!\! \left\{ \left[n_q\left(\tilde{x}^{(i)}\right)\right]^q - \left[n\left(\tilde{x}^{(i)}\right)\right]\right\} \simeq \nonumber\\
&\simeq&(q-1)n\left(\tilde{x}^{(i)}\right)\left\{ \ln n\left(\tilde{x}^{(i)}\right) + \frac{1}{2}\left[ 1 + \xi n\left(\tilde{x}^{(i)}\right)  \right] \left(\tilde{x}^{(i)}\right)^2 \right\}. \label{TA-2}
\end{eqnarray}
Fig. \ref{D-TraceA} shows the dependence of the trace anomaly on the nonextensivity $q$ (left panel) and chemical potential $\mu$ (right panel). Note that for large values of the scaled temperature $\tau$ the effects caused by the nonextensivity and by the chemical potential gradually vanish.
\begin{figure}[h]
\resizebox{0.5\textwidth}{!}{%
  \includegraphics{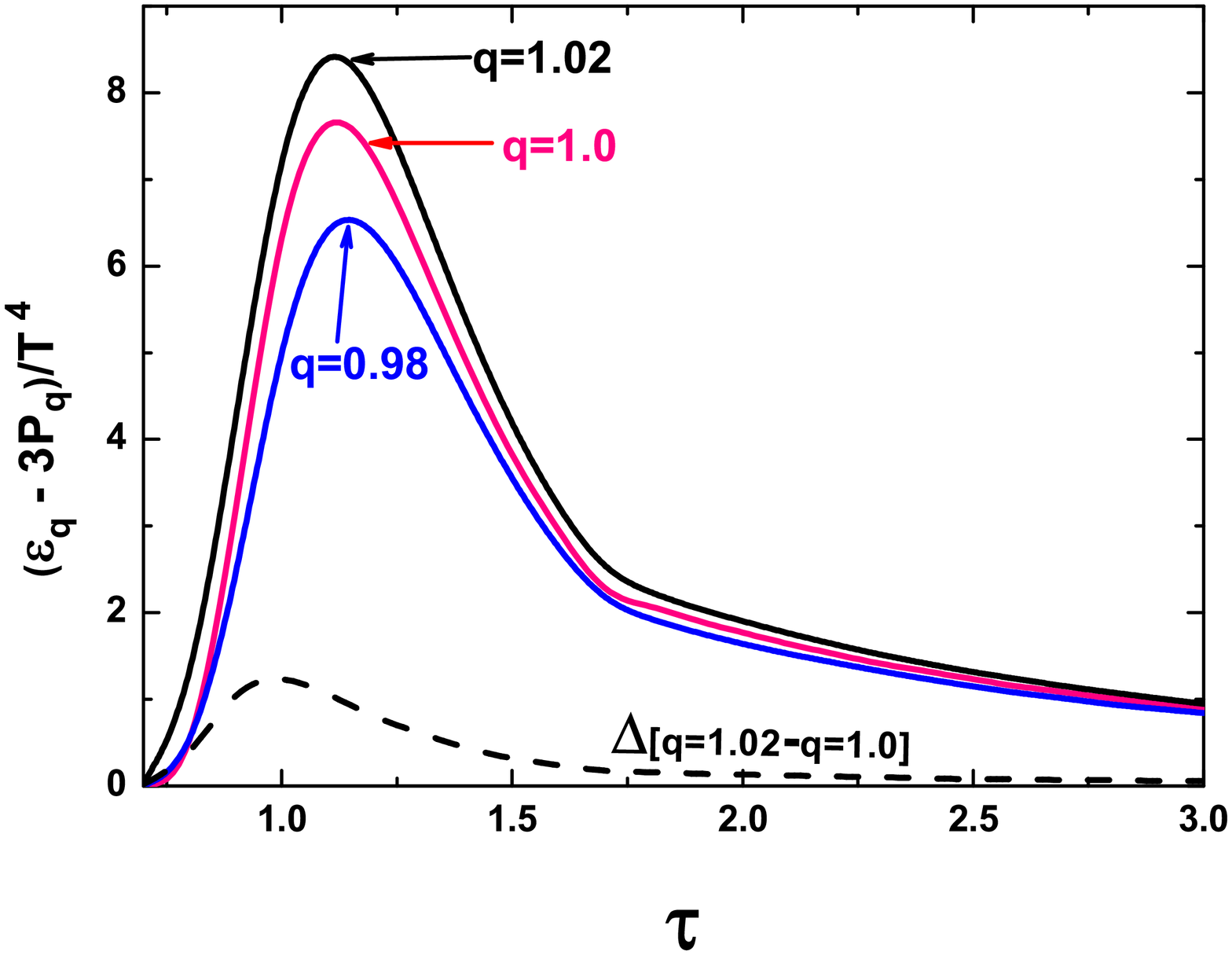}
}
\resizebox{0.51\textwidth}{!}{%
  \includegraphics{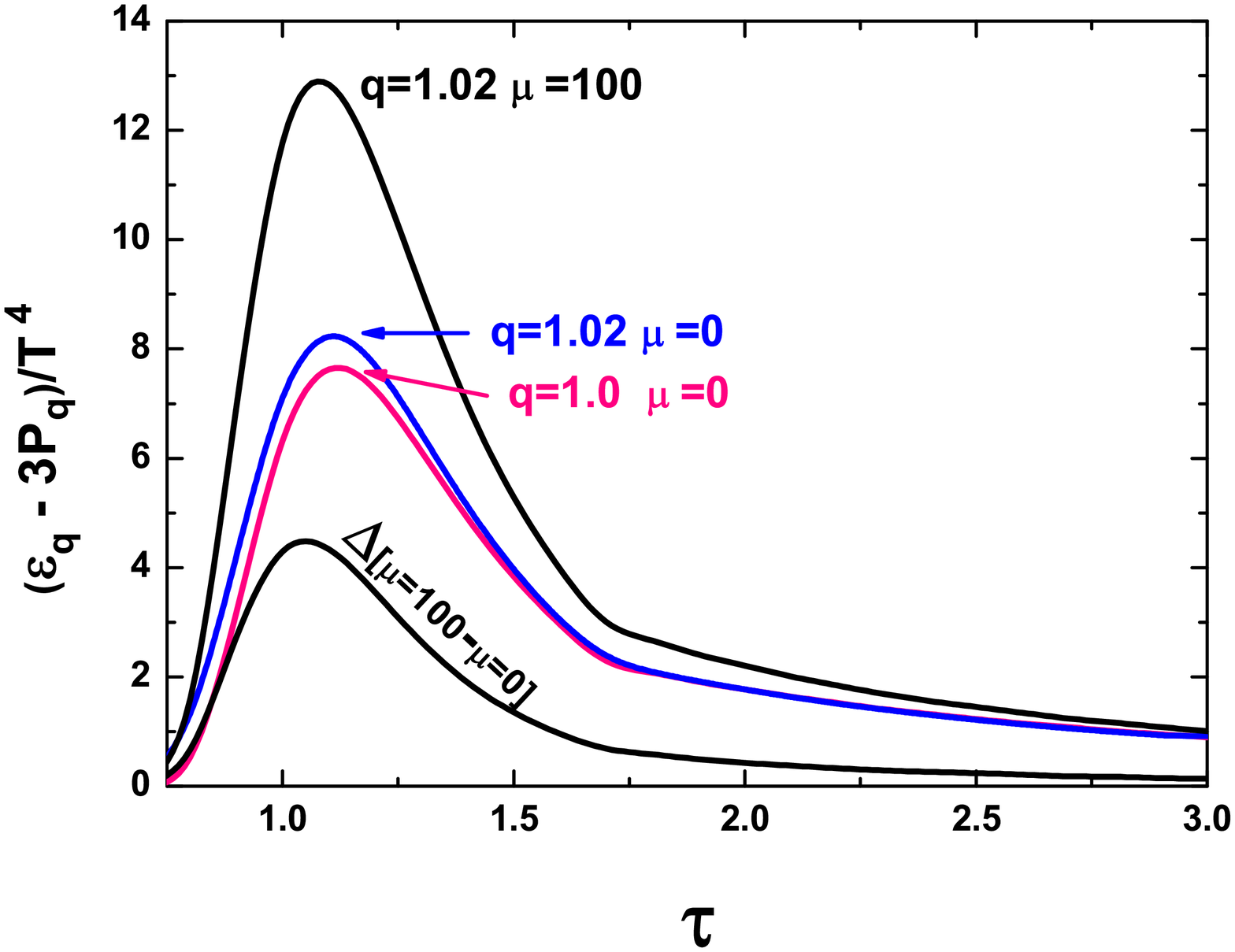}
}
\caption{(Color online) The behaviour of the change in the trace anomaly in the $qz$-QPM with $z_q=z_{q=1}$ as a function of $\tau $ for some selected values of $q$ (left panel) and the same but for some selected values of the chemical potential $\mu$ (right panel).}
\label{D-TraceA}
\end{figure}

\begin{figure}
\resizebox{0.5\textwidth}{!}{%
  \includegraphics{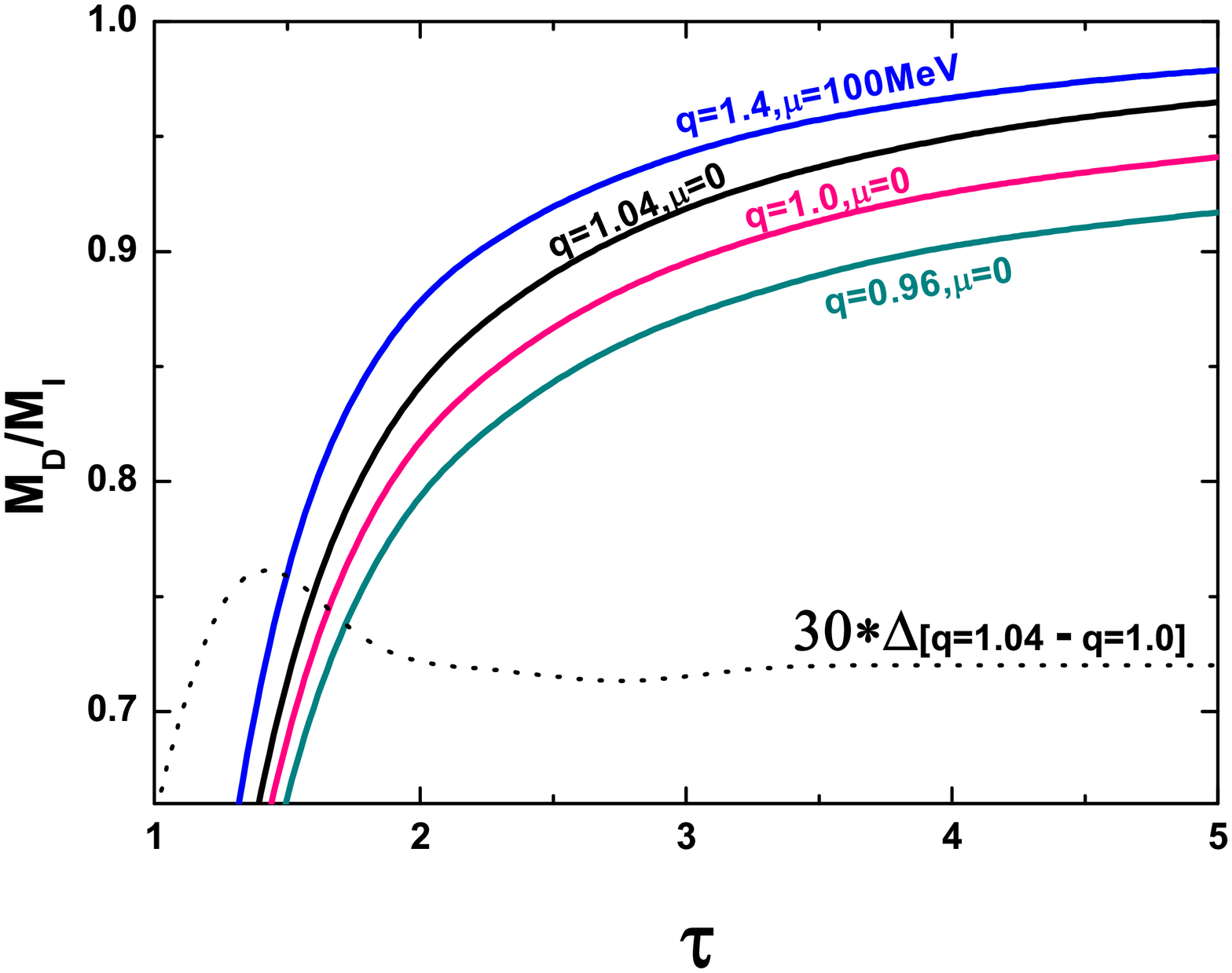}
}
\resizebox{0.51\textwidth}{!}{%
  \includegraphics{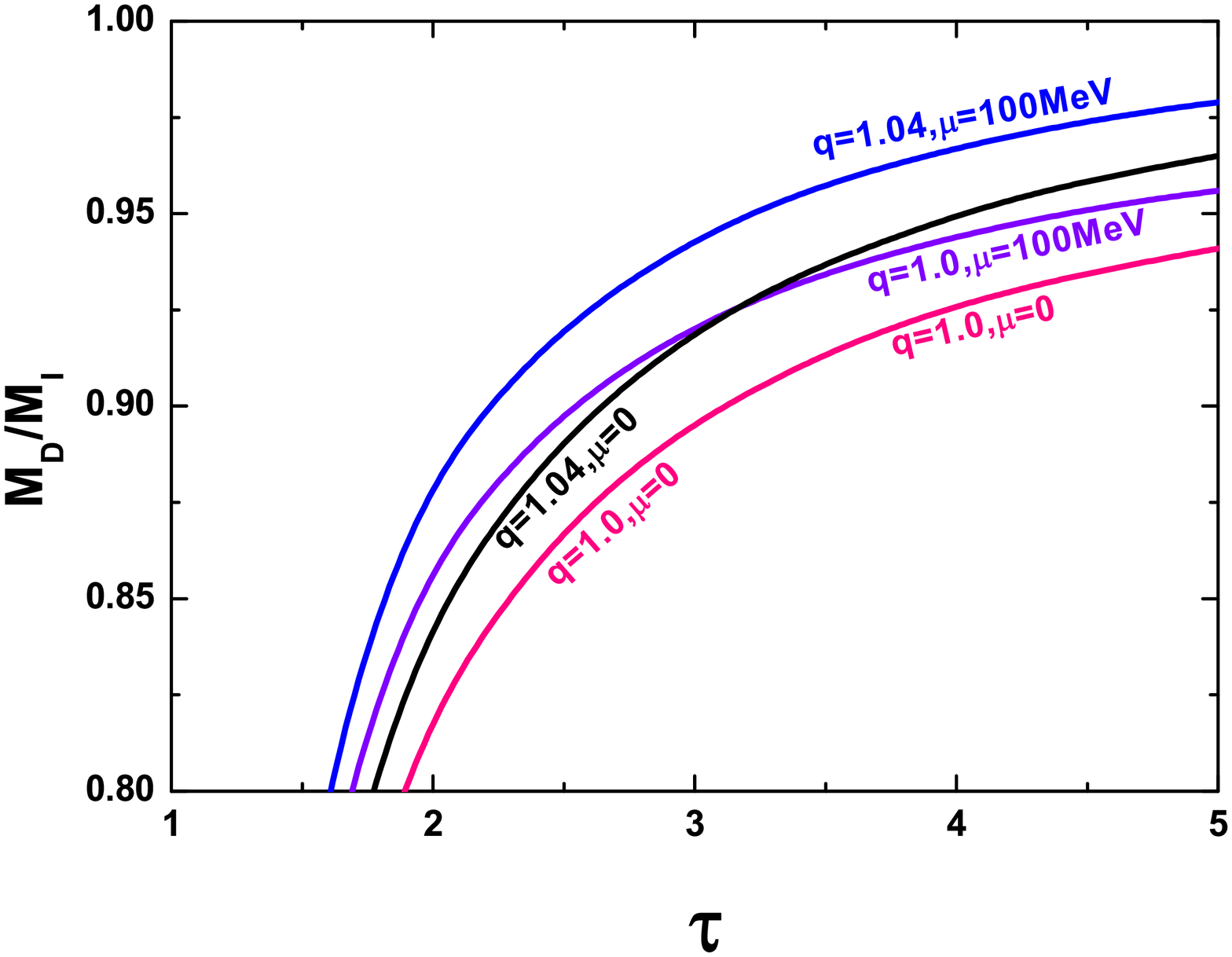}
}
\caption{(Color online) The behaviour of the ratio $M_D/M_D^I$ of the Debye masses (as defined by Eqs. (\ref{MD})) and (\ref{MDI}) as a function of $\tau$ for some selected values of $q$ (left panel) and the same but for some selected values of the chemical potential $\mu$ (right panel). The dynamics is the same as in the $z$-QPM, i.e., $z_q = z_{q=1} = z$.}
\label{f-MD-1}
\end{figure}
In Fig. \ref{f-MD-1} we present the $\tau$ dependence of the ratio $M_D/M_D^I$ of the Debye masses (as defined by Eqs. (\ref{MD})) and (\ref{MDI}) for different nonextensivities $q$ (left panel) and chemical potentials $\mu$ (right panel). Unlike the results presented in Fig. \ref{f-MD}, this time they are  caused solely by the action of the nonextensive environment with $z_q = z_{q=1} = z$ as used in the $z$-QPM.

\section{Summary and conclusions}
\label{sec:Summary}

In this work we investigate the interrelation between nonextensive statistics and the effects of dynamics in dense QCD matter. It continues our previous analysis of this problem on the example of the nonextensive version of the NJL model, the $q$-NJL. However, the complexity of its dynamics does not allow for a clear separation of the purely dynamical effects from the  nonextensive ones. Therefore, in this work, following some specific quasi-particle models ($z$-QPM) \cite{zQPM5,zQPM6,zQPM6a,zQPM11a,zQPM12,zQPM14}, we used simplified dynamics reduced to a number of well defined parameters, the effective fugacities, $z \in (0,1)$.  In this kind of QPM the masses of the quasi-particles are not modified by the interaction,
which enables the problems and inconsistencies encountered in other approaches to be avoided. The fugacities $z$ increase with temperature $T$ from very small values in the vicinity of the critical temperature, $T_{cr}$ (which corresponds to strong interactions between quarks and gluons), towards unity (which corresponds to a free gas of quarks and gluons). They modify only the argument of the exponent in the corresponding Bose-Einstein or Fermi-Dirac distributions:  $e(x)\to e(x - \ln z)$. The action of nonextensivity is different, it changes the functional form of the exponent, $e(x) \to e_q(x)$, leaving the argument $x$ unchanged. This means that the actions of nonextensivity and dynamics are complementary and cannot be replaced by each other (although sometimes they describe the same, or comparable, situations). The fugacity $z$ therefore models phenomenologically the dynamics of the mean field theory in the extensive environment and does not account for intrinsic correlations and fluctuations present in the system, while these are most naturally described phenomenologically by the nonextensivity $q$. Phenomenologically, both approaches nicely complement each other in what concerns the description of the dense QCD system. If we wanted to replace the action of nonextensivity, $q$, by the respective action of dynamics, $z$, (or vice versa) then either $z$ or $q$ would have to acquire energy dependence, which we consider as untenable.

Note that, contrary to the $q$-NJL model \cite{JRGW}, the $qz$-QPM model is formulated in such a way as to reproduce the effective fugacities of the original $z$-QPM \cite{zQPM5}, which, in turn, describes the lattice QCD results \cite{Latt-mu1,Latt-mu3}. This means that the $qz$-QPM also describes them; in fact they serve as a kind of experimental data.
Such constraints were not present in the $q$-NJL model. Therefore, our conclusions are more reliable than those presented in \cite{JRGW}. The interplay between dynamics and nonextensivity is best seen in Fig. \ref{mu-density} (left panel) which shows results for the relative densities, $R_{\rho} = \rho_q/\rho_{q=1}$, in the nonextensive environment. For $q<1$  (which corresponds to a lowering of the entropy) one observes $R_{\rho} > 1$, which can be interpreted as caused by some positive (attractive) correlations in the system and may be connected with a tighter packing of quarks. The opposite is observed for $q>1$ (corresponding to an increase of the entropy) where $R_{\rho} <1$. This can be interpreted as resulting from the repulsion of the quarks and fluctuations developing in the system. Both of these correlations and fluctuations are imposed on the effects of the interaction described by the fugacities $z_q$. This is the clearest example of dynamical effects introduced by the nonextensive environment and characterized by the nonextensivity parameter $q$.

Let us now look more closely at the results on $z_q(\tau)$ presented in Fig. \ref{zqg-1}. Note that for $q<1$, when, according to the left panel of Fig. \ref{mu-density} our system becomes more dense, one observes that $\delta z_q = z_{q<1} - z_{q=1} > 0$ and  increases with $|q-1|$ (i.e., increases with density). This means that the interaction represented by $z_q$ becomes weaker. As a result, the upper limit of $z_q=1$ (corresponding to a noninteracting gas of quarks and gluons) is reached for smaller temperature $T$,  the more so the bigger $|q-1|$ (i.e., the smaller $q$). This means that to obtain the same pressure in the system one needs a weaker interaction described by fugacity; the increasing part of it is caused by the effect of the nonextensivity $q$. In other words: the change of statistics from extensive ($q=1$) to nonextensive with $q<1$, allows the attainment of the limit of the ideal gas with weaker correlations between quarks and gluons caused by the fugacity $z$. For the $q>1$ case our system becomes, according to the left-panel of Fig. \ref{mu-density}, less dense; the correction term  needed to obtain the same pressure as in the extensive case is now negative, $\delta z_q = z_{q>1} - z_{q=1} < 0$, and $|\delta z_q|$ grows only very slowly with increasing $q$ (i.e., with decreasing density) becoming constant for higher $T$; the limit $z_q =1$ is never reached for finite temperature $T$. This is because for $q > 1$ one expects some intrinsic fluctuations (for example temperature $T$ fluctuations) which work against the dynamical interactions represented by $z$. Therefore, these interactions cannot cease and $z_q$ cannot grow too fast. In fact, with increasing $T$ they seem to become constant and one observes a kind of equilibrium between dynamics and nonextensivity.

Because the $z$-QPM \cite{zQPM5,zQPM6,zQPM6a,zQPM11a,zQPM12,zQPM14} uses the lattice QCD results \cite{LQCD1,LQCD3,LQCD4} as its input and because there are problems with nonzero chemical potential $\mu$ in the lattice calculations \cite{Latt-mu1,Latt-mu3}, the $z$-QPM was initially formulated for zero chemical potential, $\mu = 0$, which substantially limits its applications. However, anticipating the possibility of the emergence of some new lattice QCD results with the chemical potential included (if only partially), starting from \cite{zQPM12} some small amount of non-vanishing $\mu$ in the matter sector was introduced. We have therefore also allowed for some nonvanishing $\mu$. In Fig. \ref{zqmu} we show how nonzero $\mu$ influences the extracted $z_q$ for $q<1$ and $q>1$. Fig. \ref{mu-density} (right panel) shows that the relative density (both for $q<1$ and $q>1$) increases (almost) linearly with the chemical potential. Note that the possible introduction of the chemical potential in the lattice QCD calculations will change profoundly the $z$-QPM (and the $qz$-QPM); it will therefore become a third phenomenological parameter modelling the interaction. Our results shows in what direction these changes will proceed and in Appendix \ref{sec:C} we provide a scheme of expansion of the pressure in the chemical potential to allow for the possible further application of our $qz$-QPM should similar results occur in the lattice calculations \cite{Latt-mu1,Latt-mu3}.

Fig. \ref{f-MD} presents the results for the Debye mass in a nonextensive environment. Note that because it is essentially a combination of densities of quarks and gluons the results therefore resemble those for the effective fugacities. Calculations of more involved quantities, like, for example, dissipative effects would be much more involved because they would demand the use of the nonextensive version of the transport or hydrodynamic equations, which is beyond the scope of this work and will be presented elsewhere. It would also be desirable to be able to compare directly the results of $qz$-QPM with some future nonextensive lattice QCD simulations, which seems to be gaining some interest recently \cite{qLatt,qLatt1}.

Finally, Figs. \ref{Original} - \ref{f-MD-1} present some selected results on the, respectively, relative pressure and density, trace anomaly and Debye mass obtained when we use the $qz$-QPM with the same effective fugacities $z^{(i)}$ as in the $z$-QMP and calculate changes in densities and pressure induced only by changes in the nonextensivity $q$. Because in this case the original dynamics represented by $z^{(i)}$ remains intact, all changes in the results are caused only by the nonextensivity, i.e., by the fact that $|q-1| \neq 0$. These results correspond, in a sense, to the results obtained in our $q$-NJL model \cite{JRGW}, with the proviso that now our investigations are limited to $T$ above the critical temperature $T_c$ (i.e., to $\tau = T/T_c >1$ because only such are considered in $z$-QMP). In both models we observe similar dependencies of the pressure and density on the nonextensivity parameter $q$ while maintaining all dynamical parameters for a given temperature $T$ the same. They are reduced for $q<1$ and enhanced for $q>1$. This means that when changing the amount of nonextensivity one cannot keep the same pressure $P$ in the system without changing the dynamical parameters (or their temperature dependencies). Our $qz$-QPM  is therefore a simple example of such changes needed to achieve equalization of the pressure in extensive and nonextensive systems. Note that now, as a result of the pressure equalization in extensive and nonextensive systems, the relative densities, $R_{\rho} = \rho_q/\rho_{q=1}$, change with $q$ in opposite ways, becoming higher for $q<1$ and lower for $q<1$ than the density in an extensive system. This observation could be important for some new version of the $q$-NJL model, in which one could insist on keeping the same pressure for different nonextensivities and looking for the corresponding changes in its dynamical parameters (which would become $q$-dependent). Such an approach could have its further application in investigations of the EoS of dense matter.

\acknowledgments{Acknowledgments}

This research  was supported in part (GW) by the National Science Center (NCN) under contract 2016/22/M/ST2/00176.  We would like to thank warmly Dr Nicholas Keeley for reading the manuscript.

\authorcontributions{Author Contributions}
Conceptualization, Jacek Ro\.zynek and Grzegorz Wilk; Formal analysis, Jacek Ro\.zynek and Grzegorz Wilk; Software, Jacek Ro\.zynek and Grzegorz Wilk; Writing - original draft, Jacek Ro\.zynek and Grzegorz Wilk.


\conflictofinterests{Conflicts of Interest}

The authors declare no conflict of interest.

\appendix

\section{Limitations of the allowed phase space in the nonextensive approach}
\label{sec:0}

The functions $\Theta(p;i)$ ($i=g$ for gluons, $i=q$ for light quarks and $q=s$ for strange quarks) provide the limitations of the allowed phase space resulting from the condition
\begin{equation}
[1 + (q-1)x] \geq 0, \label{def}
\end{equation}
In the $q<1$ case, for gluons (with zero mass and zero chemical potential) we have that
\begin{equation}
\tilde{x}^{(i)} < \frac{1}{1-q}\qquad \Longrightarrow \qquad \frac{p}{T} < \frac{1}{1-q} + \ln z_q^{(g)}.\label{qle1}
\end{equation}
Because $z_q^{(g)} < 1$, our integral is non-vanishing (i.e., $p>0$) only for
\begin{equation}
\frac{1}{1-q} + \ln z_q^{(g)} > 0 \quad \Longrightarrow  \quad z_q^{(g)} > e\left( - \frac{1}{1-q}\right). \label{z-qle1}
\end{equation}
Stronger interactions (corresponding to smaller values of the fugacity) are in this case not allowed for the $q$ used here.
In the case of quarks (with chemical potential $\mu >0$ and with mass $m$ for strange quarks) condition (\ref{qle1}) results in the following limitation
\begin{equation}
\frac{\sqrt{p^2 + m^2}}{T} < \frac{1}{1-q} + \frac{\mu}{T} +  \ln z_q^{(i)},\quad( i=q,s). \label{pq-gle1}
\end{equation}
Now the phase space is open if:
\begin{equation}
\frac{1}{1-q} + \frac{(\mu \pm m)}{T} + \ln z_q^{(i)} > 0\qquad{\rm or}\qquad
\frac{1}{1-q} + \frac{(\mu \pm m)}{T} + \ln z_q^{(i)} < 0. \label{sol-plus-minus}
\end{equation}
In the first case the $(\mu - m)$ choice is more restrictive and results in the condition that
\begin{equation}
z_q^{(i)} > e\left( - \frac{1}{1-q}\right)\cdot e\left( - \frac{\mu - m}{T}\right),\label{q-cond1}
\end{equation}
which for $\mu =0$ and $m=0$ coincides with the corresponding condition for gluons. In the second case the choice $(\mu + m)$ is the more restrictive, for which
\begin{equation}
z_q^{(i)} > e\left( - \frac{1}{1-q}\right)\cdot e\left( - \frac{\mu + m}{T}\right) .\label{q-cond2}
\end{equation}
For nonzero mass (strange quarks) it is more restrictive than condition (\ref{q-cond1}).

In the case of $q > 1$  we have for gluons that
\begin{equation}
x^{(i)} > - \frac{1}{q-1}\qquad \Longrightarrow\qquad \frac{p}{T} > \ln z_q^{(g)} - \frac{1}{q-1}. \label{pg-qth1}
\end{equation}
In our case it is always satisfied and there are no limitations on $z_q^{(g)}$. The same situation is now in the quark sector and there are also no limitations $z_q^{(i)}$. For gluons, which are bosons, one has an additional condition, namely
\begin{equation}
e_q(x) > 1\quad   \Longrightarrow\quad  p > T \ln z_q^{(i)}\quad {\rm for~all}\quad q. \label{pgtTz}
\end{equation}
However, for $z_g < 1$ it does not introduce any further limitations.

\section{Approximate calculation of $z_q$}
\label{sec:A}

Let us denote $z_q = z + \delta $ (where $z = z(\tau)$ are the fugacities obtained in \cite{zQPM5} from lattice QCD and $\delta = z_q - z$ is the change in fugacity emerging from the nonextensive environment). We shall now calculate $\delta/z$ for the case of small $\delta$, $|\delta /z| << 1$. We start by expanding $L_q(x)$ from Eq. (\ref{aXi}),
\begin{equation}
L_q\left( x_q;\delta \right) = \ln_{2-q}\left[ 1 - \xi e_{2-q}\left(- x_q\right)\right] = \ln_{2-q} (X),
\label{Lqdef}
\end{equation}
in $\delta$ and keeping only linear terms:
\begin{equation}
L_q\left( x_q; \delta\right) \simeq L_q \left( x_q; \delta =0\right) + \frac{\partial L_q\left( x_q; \delta \right)}{\partial \delta}\bigg |_{\delta =0}\cdot  \delta. \label{Lqapprox}
\end{equation}
Denoting
\begin{equation}
X = 1 - \xi E,\quad E = e_{2-q}\left(-x_q\right),\quad x_q = y - \ln z_q, \label{XEx}
\end{equation}
one can write that (cf., Eqs. (\ref{eqe}) and (\ref{enqd}))
\begin{eqnarray}
\frac{\partial L_q\left( x_q; \delta \right)}{\partial \delta} &=& \frac{\partial L_q(X)}{\partial X}\cdot \frac{\partial X}{\partial E}\cdot \frac{\partial E}{\partial x_q}\cdot \frac{\partial x_q}{\partial z_q}\cdot \frac{\partial z_q}{\partial \delta},\label{LEx}\\
\frac{\partial L_q(X)}{\partial X} &=& \frac{\partial \ln_{2-q}(X)}{\partial X} = X^{-q},\label{pX}\\
\frac{\partial X}{\partial E}  &=& -\xi,\qquad \frac{\partial E}{\partial x_q} = - \left[ e_{2-q}(-x_q)\right]^q,\qquad
\frac{\partial x_q}{\partial z_q} = - \frac{1}{z_q},\qquad \frac{\partial z_q}{\partial \delta} =1 \label{px}
\end{eqnarray}
obtaining (note that for $\delta = 0$ $x_q \to x - \ln z$)
\begin{equation}
\!\!\!\!\!\frac{\partial L_q\left( x_q; \delta \right)}{\partial \delta}\bigg |_{\delta =0}\! =\! - \frac{\xi}{z}\left[n_q(x;z)\right]^q \quad {\rm and}\quad L_q\left( x_q; \delta\right) \simeq \ln_{2-q}\left[ 1 - e_{2-q}\left(-x;z\right)\right] - \frac{\xi}{z} \left[ n_q\left(x;z\right)\right]^q  \delta.  \label{Resdelta}
\end{equation}
The integrals of the type presented in Eqs. ({\ref{gluons}) and (\ref{quarks}) can therefore be rewritten as integrals over
\begin{eqnarray}
\Delta L_q &=& L(x;z) - L_q(x;\delta) = \ln [ 1 - e( -x;z)] - \ln_{2-q}\left[ 1 - \xi e_{2-q}\left( -x_q; z_q\right)\right] \simeq\nonumber\\
&\simeq& \left\{ \ln [ 1 - e( -x;z)] - \ln_{2-q}\left[ 1 - \xi e_{2-q}( -x; z)\right]\right\} + \xi \left[ n_q(x;z)\right]^q\cdot \frac{\delta}{z} = \nonumber\\
&& = \left[ I(x;z)  - I_q(x;z)\right] + \xi \left[ n_q(x;z)\right]^q\cdot \frac{\delta}{z}. \label{Dz-final}
\end{eqnarray}
Because $z_q$ is obtained from the condition that $\Delta L_q = 0$, in the first approximation the correction term $\delta$ is equal to
\begin{eqnarray}
\delta &=& \xi z\cdot \frac{ \int_0^{\infty}dp p^2 \left[ I(x;z) - I_q(x;z)\Theta(p)\right]}{ \int_0^{\infty} dp p^2 \left[ n_q(x;z)\right]^q\Theta(p)}. \label{Dz-finappr}
\end{eqnarray}
$\Theta(p)$ represents the possible limitation of the phase space caused by the nonextensivity (cf. Appendix \ref{sec:0}). It depends on the nonextensivity parameter $q$ and on the type of particle considered (gluons, light quarks or strange quarks).

Formula (\ref{Dz-finappr}) can be further approximated by expanding it in $q-1$ and retaining only the linear terms in $(q-1)$. Following Eqs. (\ref{L(xi)}) and (\ref{DL(xi)}) one obtains that
\begin{equation}
I_q(x;z) \simeq I(x;z) + \frac{1}{2}(q-1) \Delta I(x;z)\quad {\rm where} \quad \Delta I(x;z) = - \ln^2[1-\xi e(-x)] + n(x;z) x^2.  \label{DeltaI}
\end{equation}
Similarly, following Eqs. (\ref{eq-qnq}) and (\ref{Deq-qnq}), one has that
\begin{equation}
n_q^q \simeq n(x;z) + (q-1)n(x;z)\Delta_q[n(x;x)]\quad {\rm where}\quad \Delta_q[n(x;z)] = \ln n(x;z) + \frac{1}{2}[1 + \xi n(x;z)] x^2. \label{eq-qnq-1}
\end{equation}
Therefore
\begin{equation}
\xi\frac{\delta}{z} = \frac{A(x;z) - (q-1)B(x;z)}{C(x;z) + (q-1)D(x;z)} \simeq \frac{A(x;z)}{C(x;z)} - (q-1) \left\{ \frac{B(x;z)}{C(x;z)} + \frac{A(x;z) D(x;z)}{[C(x;z)]^2}\right\},\label{Fires}
\end{equation}
where
\begin{eqnarray}
A(x;z) &=& \int_0^{\infty} dp p^2 I(x;z) [ 1 - \Theta(p)], \label{A}\\
B(x;z) &=& \frac{\xi}{2} \int_0^{\infty} dp p^2 \Delta I(x;z) \Theta(p), \label{B}\\
C(x;z) &=& \int_0^{\infty} dp p^2 n(x;z)\Theta(p), \label{D}\\
D(x;z) &=& \int_0^{\infty} dp p^2 \Theta(p) n(x;z)\cdot \left\{ \ln n(x;z) + \frac{1}{2}[ 1 + \xi n(x;z)] x^2 \right\}.
\end{eqnarray}
In practical applications it turns out that $A(x;z) \simeq 0$, therefore
\begin{equation}
\frac{\delta}{z} \simeq  - \xi \cdot (q-1) \frac{B(x;z)}{C(x;z)} = (1 - q)\cdot \frac{\int_0^{\infty} dp p^2 \left\{ \ln^2[ 1 - \xi e(-x;z)] + n(x;z) x^2\right\}}{2\int_0^{\infty} dp p^2 n(x;z)}. \label{Final-delta}
\end{equation}

\section{Some selected first order expansions in $(q-1)$}
\label{sec:B}

List of some useful first order expansions in $q-1$ \footnote{We do not address the question of the applicability of such an approach, assuming its validity for the range of variables used here (cf. \cite{TOGW}).}.
\begin{eqnarray}
e_q(x) &=& \left[ 1 + (q-1)x\right]^{\frac{1}{q-1}} \simeq e(x) - (q-1)e(x)\Delta \left[e(x)\right],\qquad \Delta \left[e(x)\right] = \frac{1}{2}x^2. \label{eq-app}\\
e_{2-q}(-x) &\simeq&  e(-x) + (q-1)e(-x)\Delta [e(x)]\qquad ({\rm because}\quad  \Delta [e(-x)] = \Delta [e(x)]), \label{e(2-q)-app}\\
n_q(x) &=& \frac{1}{e_q(x) - \xi} \simeq \frac{1}{[e(x) -\xi] - \frac{1}{2}(q-1) e(x) x^2} \simeq  n(x) + (q-1) n(x)\Delta[n(x)], \label{eq-nq}\\
\ln n_q(x) &\simeq& \ln n(x) + (q-1)\Delta[n(x)]\qquad {\rm where}\qquad \Delta[n(x)] = \frac{1}{2}\left[ 1 + \xi n(x)\right] x^2,  \label{eq-lnnq}\\
n_q^q &=& n_q \cdot e\left[ (q-1)\ln n_q\right] \simeq n_q \left[ 1 + (q-1) \ln n_q\right] \simeq n(x) + (q-1) n(x) \Delta_q[n(x)], \label{eq-qnq}\\
&& {\rm where}\quad \Delta_q[n(x)] = \ln n(x) + \Delta[n(x)]. \label{Deq-qnq}
\end{eqnarray}
More involved expressions.
\begin{eqnarray}
\left[1-\xi e_{2-q}(-x)\right] &\simeq& [1 - \xi e(-x)] - (q-1) \xi e(-x) \Delta [e(x)],  \label{1-xieq}\\
\ln\left[ 1 - \xi e_{2-q}(-x)\right] &\simeq&   \ln[1\! -\! \xi e(-x)] + (q-1)\left[ \frac{\xi e(-x)}{1\! -\! \xi e(-x)}\right] \Delta[e(x)]. \label{ln(1-xiq)}
\end{eqnarray}

The corresponding $q$-logarithm and $(2-q)$-logarithm functions to be used in what follows are connected with the $q$-exponential function $e_q(x)$ and its dual $e_{2-q}(x)$:
\begin{eqnarray}
\ln_q X = \frac{X^{q-1} - 1}{q - 1} \stackrel{q \rightarrow 1}{\Longrightarrow} \ln X\qquad &{\rm and}&\qquad \ln_{2-q}X = \frac{X^{1-q}-1}{1-q} \stackrel{q \rightarrow 1}{\Longrightarrow} \ln X,  \label{lnq}\\
\ln_q \left[ e_q(X) \right] = X,\qquad &{\rm and}& \qquad\ln_{2-q} \left[ e_{2-q}(X) \right] = X ,\label{qlne}\\
\ln_{2-q}X &=& - \ln_q\left( \frac{1}{X}\right). \label{LvsL}
\end{eqnarray}

From them one gets that:
\begin{eqnarray}
\ln_q X &=& \frac{X^{q-1} - 1}{q-1} = \frac{e^{(q-1)\ln X} - 1}{q-1}\simeq \ln X + \frac{1}{2} (q-1) \ln^2 X. \label{lq-app}\\
\ln_{2-q} X &=& \frac{ X^{1-q} - 1}{1-q} \simeq \ln X + \frac{1}{2} (1-q)\ln^2 X \label{l(2-q)-app}\\
L^{(\xi)}_{2-q}(x) &=& \ln_{2-q} \left[ 1 - \xi e_{2-q}(-x)\right] \simeq \ln\left[ 1 - \xi e_{2-q}(-x)\right] + \frac{1}{2}(1-q)\ln^2\left[ 1 - \xi e_{2-q}(-x)\right] \simeq \nonumber\\
&\simeq& \, \ln[1 - \xi e(-x)] + \frac{1}{2}(q-1)\Delta[\ln(x)]; \label{L(xi)}\\
\Delta[\ln(x)] &=& - \ln^2[1-\xi e(-x)] + \left[ \frac{e(-x)}{1-\xi e(-x)}\right] x^2 = - \ln^2[1-\xi e(-x)] + n(x;z) x^2.   \label{DL(xi)}
\end{eqnarray}

Finally, the generalization of the relation $n_q(x) + n_{2-q}(-x) = 1$ to the case where the effective particle densities are given not by $n_q$ but by $n_q^q$ is approximately given by
\begin{eqnarray}
n_q^q(x) + n_{2-q}^{2-q}(-x)\! &=&\! 1 + (q-1)\left\{ n(x) \Delta_q[n(x)] - n(-x)\Delta_{2-q}[n(-x)]\right\} \simeq \label{Ua}\\
\!&\simeq&\! 1 + (q-1)\left\{ n(x)\ln n(x) - n(-x)\ln n(-x) + \frac{1}{2}(1 + \xi)x^2[ n(x) - n(-x)]\right\}. \nonumber
\end{eqnarray}

\section{Expansion of pressure in chemical potential $\mu $}
\label{sec:C}

In the case when we allow for a chemical potential $\mu$, in some applications we need to know the expansion of the pressure $P$ (as given by Eq. (\ref{P})) in the chemical potential $\mu$ (in fact in $\tilde{\mu} = \mu/T = \beta \mu < 1$). We present below the two first terms of such an expansion,
\begin{equation}
P_q = \frac{1}{\beta V} \ln_q\left( \Xi_q\right) \simeq \frac{1}{\beta V}\left\{  \ln_q\left( \Xi_q\right)\bigg |_{\mu =0} + \frac{\partial \ln_q\left( \Xi_q\right)}
{\partial \mu}\bigg |_{\mu =0}\cdot \mu + \frac{1}{2}\frac{\partial^2 \ln_q\left( \Xi_q\right)}
{\partial \mu^2}\bigg |_{\mu =0}\cdot \mu^2 \right\} \label{Nnq}
\end{equation}
where $\ln_q\left( \Xi_q\right)$  is given by Eq. (\ref{aXi}) and
\begin{eqnarray}
\frac{\partial \ln_q\left( \Xi_q\right)}{\partial \mu} \bigg |_{\mu =0} &=& \int \frac{d^3 p}{(2\pi)^3}\sum_i \nu_i
\left\{ n_q\left[\tilde{x}^{(i)}_q(\mu=0)\right]\right\}^q = \sum_i \nu_i \rho_q^{(i)}(\mu=0), \label{D1}\\
\frac{\partial^2 \ln_q\left( \Xi_q\right)}{\partial \mu^2} \bigg |_{\mu =0} &=& \int \frac{d^3 p}{(2\pi)^3}\sum_i \nu_i
\frac{\partial}{\partial \mu}\left\{ n_q\left[\tilde{x}^{(i)}_q(\mu)\right]\right\}^q \bigg |_{\mu =0} = \nonumber\\
&=& \beta q \int \frac{d^3 p}{(2\pi)^3}\sum_i \nu_i \left\{ n_q\left[\tilde{x}^{(i)}_q(\mu=0)\right]\right\}^{q+1}\cdot \left\{e_q\left[ \tilde{x}^{(i)}_q(\mu=0)\right]\right\}^{2-q}. \label{D2}
\end{eqnarray}
We have used here Eq. (\ref{density}) and Eqs. (\ref{Nnq}) - (\ref{D1-xmu}) (with $n_q\left[ \tilde{x}_q(\mu)\right]$, $e_q\left[ \tilde{x}_q(\mu)\right]$ and $x_q(\mu)$ are defined by Eqs. (\ref{eqe}) and (\ref{enqd})) and $\left\{ n_q\left[\tilde{x}_q(\mu)\right]\right\}^q = N_q(\mu)$):
\begin{eqnarray}
\frac{\partial N_q}{\partial \mu} &=& \frac{\partial N_q}{\partial n_q}\cdot\frac{\partial n_q}{\partial e_q}\cdot\frac{\partial e_q}{\partial \tilde{x}_q}\cdot\frac{\partial \tilde{x}_q}{\partial \mu} = \beta q n_q^{q+1}\cdot e_q^{2-q}, \label{D1}\\
\frac{\partial N_q}{\partial n_q} &=& q n_q^{q-1},\qquad\qquad\qquad\qquad\qquad\qquad \frac{\partial n_q}{\partial e_q} = - \left( e_q - \xi\right)^{-2} = - n_q^2,\label{1-2_dernEa}\\
\frac{\partial e_q}{\partial \tilde{x}_q} &=& \left[1 + (q-1)\tilde{x}_q\right]^{\frac{2-q}{q-1}} = e_q^{2-q},  \label{1-2_derExa}\qquad\qquad
\frac{\partial \tilde{x}_q}{\partial \mu} = - \beta. \label{D1-xmu}
\end{eqnarray}

\bibliographystyle{mdpi}
\makeatletter
\renewcommand\@biblabel[1]{#1. }
\makeatother

\end{document}